\documentclass[prl,twocolumn,showpacs,amsmath,amssymb,superscriptaddress,floatfix,reprint,nofootinbib]{revtex4-2}
\usepackage{amssymb}
\usepackage{amsbsy}
\usepackage{amsmath}
\usepackage{mathrsfs}
\usepackage{epsfig}
\usepackage{graphicx}
\usepackage{array}
\usepackage{textcomp}
\usepackage{xcolor}
\usepackage{braket}
\usepackage{hhline}
\usepackage{dsfont}
\usepackage{comment}
\usepackage{bm,hyperref}
\usepackage[justification=raggedright,font=small]{caption}
\usepackage[titletoc,toc,title]{appendix}
\usepackage[normalem]{ulem}

\newcommand{\be}{\begin{equation}}
\newcommand{\ee}{\end{equation}}
\newcommand{\iea}{\begin{equation}\begin{aligned}}
\newcommand{\fea}{\end{aligned}\end{equation}}
\newcommand{\n}{\nonumber \\ }

\newcommand{\mbZ}{\mathbb{Z}}
\newcommand{\mH}{\mathcal{H}}
\newcommand{\hT}{\hat{T}}
\newcommand{\hQ}{\hat{Q}}
\newcommand{\hh}{\hat{h}}
\newcommand{\half}{\frac{1}{2}}
\newcommand{\hD}{\hat{D}}
\newcommand{\TA}{\tilde{A}}

\begin{document}

\author{Fiona J. Burnell}
\affiliation{School of Physics and Astronomy, University of Minnesota, Minneapolis, Minnesota 55455, USA}
\author{Sanjay Moudgalya}
\affiliation{Department of Physics and Institute for Quantum Information and Matter,
California Institute of Technology, Pasadena, California 91125, USA}
\affiliation{Walter Burke Institute for Theoretical Physics, California Institute of Technology, Pasadena, California 91125, USA}
\affiliation{Department of Physics, Technische Universit\"{a}t M\"{u}nchen (TUM), James-Franck-Str. 1, 85748 Garching, Germany}
\affiliation{Munich Center for Quantum Science and Technology (MCQST), Schellingstr. 4, 80799 M\"{u}nchen, Germany}
\author{Abhinav Prem}
\affiliation{School of Natural Sciences, Institute for Advanced Study, Princeton, New Jersey 08540, USA}

\title{Filling constraints on translation invariant dipole conserving systems}
\date{\today}

\begin{abstract}
Systems with conserved dipole moment have drawn considerable interest in light of their realization in recent experiments on tilted optical lattices.
An important question for such systems is delineating the conditions under which they admit a unique gapped ground state that is consistent with all symmetries.
Here, we study one-dimensional translation-invariant lattices that conserve U$(1)$ charge and $\mathbb{Z}_L$ dipole moment, where discreteness of the dipole symmetry is enforced by periodic boundary conditions, with $L$ the system size.
We show that in these systems, a symmetric, gapped, and non-degenerate ground state requires not only integer charge filling, but also a fixed value of the dipole filling, while other fractional dipole fillings enforce either a gapless or symmetry-breaking ground state.
In contrast with prior results in the literature, we find that the dipole filling constraint depends both on the charge filling as well as the system size, emphasizing the subtle interplay of dipole symmetry with boundary conditions.
We support our results with numerical simulations and exact results.
\end{abstract}

\maketitle

%%%%%%%%%%%%%%%%%%%%%%%%%%%%%%%%%

\paragraph{Introduction:} Quantum many-body systems that conserve charge as well as higher multipole moments are an exciting frontier in the study of strongly correlated quantum matter. While the multipolar (or polynomial shift) symmetries that result in higher moment conservation laws recently elicited interest given their close kinship with ``fractonic" phases~\cite{pretko2018gauge,gromov2019multipole}, the intriguing phenomenology that results from such symmetries has endowed them with a life of their own in condensed matter, high energy, and quantum information~\cite{pretko2018elastic,gromov2019chiral,gorantla2022oned,gorantla2022twod,leo2022lifshitz,gorantla2022graph,brauner2023}.
Recent experimental realizations of dipole conserving dynamics (up to exponentially long prethermal timescales)~\cite{bakr2020subdiff,scherg2021tilt, kohlert2021experimental, zahn2022driven} has further fueled the burgeoning interest in understanding the ground state(s) of dipole conserving quantum systems~\cite{prem2017emergent,yuan2020fsf,chen2021fsf,dubinkin2021dipole,stahl2022,kapustin2022,lake2022dbhm,lake2022luttinger,zechmann2022dbhm}.  

In dipole conserving systems, single particle hopping is forbidden, and dynamics occurs through correlated hopping processes involving two (or more) particles.
This leads to striking dynamical behavior, including anomalously slow dynamics~\cite{chamon2005,prem2017glassy,pai2019localize,knap2021slow,hahn2021hsf}, Hilbert space fragmentation~\cite{sala2019ergodicity,khemani2020hilbert,rakovszky2019statistical, moudgalya2019thermalization, morningstar2020freeze, moudgalya2021hilbert,feng2022hilbert}, and unconventional hydrodynamic universality classes~\cite{gromov2020hydro,feldmeier2020subdiff,iaconis2021subdiff,grosvenor2021hyrdo,moudgalya2021spectral,glorioso2022hydro,qi2023hydro}.
However, the inherently interacting nature of correlated hopping makes determining the ground state physics of \textit{generic} dipole conserving systems challenging.
Further complicating matters, many 1d dipole conserving models studied have extra unconventional symmetries (due to fragmentation~\cite{rakovszky2019statistical, moudgalya2021hilbert}), that can significantly alter the nature of the ground states~\cite{rakovszky2019statistical,moudgalya2019thermalization}.
It is thus desirable to obtain non-perturbative results constraining the low-energy spectrum, which rely only on the manner in which microscopic symmetries act on the many-body Hilbert space and not on specifics of any Hamiltonian.
One such result is the Lieb-Schultz-Mattis (LSM) theorem~\cite{LSM1961} which, along with its generalizations~\cite{oshikawa2,hastings2004,hastings2005,sims2007lsm},provides an example of minimal microscopic data constraining the universal long-wavelength physics of lattice spin systems.
LSM-type theorems now exist for various internal and spatial symmetries~\cite{oshikawasenthil,degfordummies,roy2012,sid2013,zaletel2015,watanabe2015,watanabe2016,hsieh2016,lu2017b,furuya2017,po2017,bultinck2018,watanabe2018,ryuLSM,else2021qc, aksoy2021lsm, aksoy2023lsm}; the role of spatial symmetries in these theorems underlies their deep connection with crystalline symmetry protected topological (cSPT) phases~\cite{cheng2016set,cho2017anomaly,lu2017,huang2017,qi2017,yang2018,jian2018,metlitski2018,shiozaki2018,cheng2019fermionic,jiang2019generalized,yao2019,else2020lsm,chen2021lsm}.
For example, in charge-conserving translation invariant 1d systems, a gapped, symmetric, and non-degenerate ground state is permitted only when the filling per unit cell is an integer\footnote{We emphasize that the filling constraint is not an anomaly~\cite{seiberg2022lsm}.}, which labels distinct ``weak" SPT phases that are distinct from the trivial phase only in the presence of translation symmetry~\cite{cheng2016set}. 
Establishing similar constraints is subtle for multipole symmetries, which intertwine non-trivially with spatial symmetries, and \textit{cannot} be considered internal symmetries~\cite{gromov2019multipole}.
Crucially, the symmetry itself depends sensitively on boundary conditions: in 1d translation invariant systems, dipole symmetry is a continuous $\text{U}(1)$ symmetry on an infinite lattice but a discrete $\mbZ_L$ symmetry on a closed $L$-site ring~\cite{gorantla2022oned}, a distinction which was ignored in earlier considerations of LSM constraints for dipole symmetry~\cite{prem2020lsm,dubinkin2021lsm}.\footnote{This subtlety is absent for subsystem symmetries, for which Refs.~\cite{prem2020lsm,dubinkin2021lsm} also derived LSM constraints.}
Here, we show that this distinction must be carefully accounted for, and we obtain a non-perturbative filling constraint on the ground state(s) of 1d dipole conserving, translation invariant systems with periodic boundary conditions (PBC). 
For bosonic/integer spin systems at integer charge filling we show that a symmetric, gapped, and non-degenerate ground state is permitted at a single dipole filling (whose precise value depends on the charge filling and system size), where in PBC the dipole filling is well-defined only modulo an integer.  
At other dipole fillings, the ground state must be either gapless or symmetry-breaking.
We establish similar constraints for fermionic/half-integer spin systems, and support our claims through exact solutions and numerical analysis.

%%%%%%%%%%%%%%%%%%%%%%%%%%%%%%%%%

\paragraph{Global symmetries:} We consider arbitrary 1d lattice systems with local interactions invariant under the following symmetries: $\mathbb{Z}_L^T$ lattice translation, $\text{U}(1)_c$ charge conservation, and $\mathbb{Z}_L^D$ \textit{discrete} dipole symmetry.
The dipole symmetry is forced to be discrete here since we consider a finite lattice with $L$ sites (where $L$ is arbitrarily large) with PBC, in contrast to previous works that treat it as a continuous $\text{U(1)}$ symmetry.
The operator $\hQ = \sum_{j=1}^L \hat{n}_j$ measures the total $\text{U}(1)_c$ charge, where $\hat{n}_j = \hat{b}^\dagger_{j} \hat{b}_{j}$ is the on-site number operator and the creation/annihilation operators $\hat{b}^\dagger_{j}, \hat{b}_{j}$ obey canonical commutation (anti-commutation) relations for bosonic (fermionic) systems. 
For spin-$s$ degrees of freedom (d.o.f.'s), $\hQ = \sum_{j=1}^L{\hat{S}_j^z}$, where $\hat{\vec{S}}_j$ is the spin-$s$ operator at site $j$.
%
%We restrict to a single site per unit cell and one d.o.f. per site for simplicity.
%
The global $\text{U}(1)_c$ symmetry is generated by $e^{i \alpha \hQ}$, which sends $\hat{b}_{j} \to e^{i \alpha} \hat{b}_j$ under conjugation; $\alpha \in [0, 2\pi)$ is a continuous periodic variable due to charge conservation.
$\hT = e^{i a \hat{P}}$ is the generator of translation symmetry, which conjugates $\hat{b}_j \to \hat{b}_{j+1}$, where $\hat{P}$ is the total momentum operator and $a$ is the lattice constant (which we set equal to 1). 
On the translation invariant infinite chain, the dipole operator is defined as $\hQ_D = \sum_{j=1}^L j \hat{n}_j$, leading to the U$(1)$ dipole symmetry considered in Refs.~\cite{prem2020lsm,dubinkin2021lsm} i.e., the symmetry of the form $e^{i \beta \hQ_D}$ with $\beta \in [0,2\pi)$ 
(analogously, for spins, $\hQ_D = \sum_{j=1}^L j S_j^z$; hereinafter we restrict to $s \in \mbZ$, with the half-integer case discussed in the SM~\cite{supmat}).
On a finite ring, however, $\hQ_D$ is incompatible with PBC, required for translation invariance, and the symmetry generator for the $\mbZ_L^D$ dipole symmetry is:
\be
\label{eq:dipolegen}
\mathbb{Z}_L^D \text{ dipole}: \hD = e^{i \frac{2\pi}{L} \hQ_D}.
\ee
Under conjugation by $\hD$, we have $\hat{b}_j \rightarrow e^{i \frac{2 \pi }{L}  j} \hat{b}_j$. 
Hence only operators $e^{i \beta \hQ_D}$ with $\beta$ a multiple of $2 \pi/L$ are symmetry operators: this constraint comes from demanding single-valuedness of this phase rotation under translation by $L$ for all U$(1)_c$ sectors; hence the ``dipole charge" $Q_D$ is well-defined only modulo $L$. 
We emphasize that the discreteness of dipole symmetry is enforced by the boundary conditions, and $\beta$ would be quantized even for a disordered dipole conserving system on a finite ring~\cite{moudgalya2019thermalization}.\footnote{Though continuous dipole symmetry could emerge in the $Q=0$ sector~\cite{dubinkin2021lsm}, we consider systems with exact dipole symmetry here.}
There is an origin-dependence of the dipole operator and its filling fraction, and here we fix it to a lattice site, which at integer charge filling fixes the dipole filling fraction modulo $1$.
Our main result, however, is independent of this choice~\cite{supmat}.
%
%%%%%%%%%%%%%%%%%%%%%%%%%%%%%%%%%

%
\paragraph{Symmetry algebra:} While the charge $\hQ$ commutes with $\hT$, the dipole $\hD$ obeys the algebra $\hT \hD = \exp(-i 2 \pi \hat{Q}/L) \hD \hT$~\cite{seidel2005incompressible, moudgalya2019thermalization}.
We restrict our attention to systems with fixed charge so that $\hat{Q}$ can be replaced by its eigenvalue $Q$; $\hD$ and $\hT$ then satisfy: 
\begin{equation}
    \hT \hD = \exp\left(-i 2 \pi \nu \right) \hD \hT \, ,
\label{eq:dipolecommutation}
\end{equation}
where $\nu = Q/L$ is the charge filling fraction.
At fixed integer filling, the $\hT$ and $\hD$ commute, and the full symmetry group is $G = \mathbb{Z}_L^{T} \times  \mathbb{Z}_L^{d} \times \text{U}(1)_c$.
In contrast, at fixed \textit{fractional} charge filling, the symmetry group is $G_{\nu} = \left(\mathbb{Z}_L^{T} \times \mathbb{Z}_L^{d}\right) \leftthreetimes_{\nu} \text{U}(1)_c$ where $\leftthreetimes_\nu$ indicates that the symmetry algebra is a central extension\footnote{Since the commutator between $\hT$ and $\hD$ formally involves an operator, the symmetry group is not centrally extended when the total charge is {\it not} fixed~\cite{gorantla2022oned}; we do not consider this here.} of $\mbZ_L^T \times \mbZ_L^D$ by the $\text{U}(1)_c$ filling fraction.
The resemblance with the magnetic translation group~\cite{zak1964a,zak1964b} in charge conserving (i.e., $\text{U}(1)_c$ symmetric) systems is not accidental; the thin-torus limit of 2d fractional quantum Hall systems results in 1d dipole conserving models~\cite{bergholtz2006one,bergholtz2008quantum,lee2015geometric, moudgalya2020quantumhall}.
A symmetric, gapped, and unique ground state is not permitted at fractional charge filling since the dipole algebra Eq.~\eqref{eq:dipolecommutation} does not admit one-dimensional representations when $\nu \notin \mathbb{Z}$.
In fact, \textit{every} translation invariant eigenstate of any dipole conserving Hamiltonian can be shown to be (at least) $q$-fold degenerate for $\nu = p/q$ ($p,q$ co-prime)~\cite{seidel2005incompressible}.
%
%For a translation invariant eigenstate $\ket{\phi}$, $\hD \hT \ket{\phi} = e^{ 2 \pi i \nu } \hT \left( \hD  \ket{\phi} \right )$, so the dipole operator shifts the many-body momentum by $2 \pi \nu$ (mod $2 \pi$).
%
%For fractional $\nu$, this results in a distinct translation eigenvalue; hence the states $\ket{\phi}$ and $\hD \ket{\phi}$ are orthogonal.
%
%Since $\hat{D}$ is a symmetry operator, these states have equal energies, which leads to a $q$-fold degeneracy through the spectrum.
%
As a corollary, we obtain the constraint that a symmetric, gapped, and non-degenerate ground state is only allowed at integer charge fillings $\nu \in \mbZ$, where for gapped states the different integer labels correspond to distinct weak SPT phases in 1D protected by $\text{U}(1)_c$ and $\mbZ_L^T$~\cite{else2021qc,cheng2016set}.
For half-integer spins, the constraint is modified: a unique, symmetric, and gapped ground state can only be obtained when $\nu \in \mbZ + 1/2$~\cite{supmat}.
Given the above charge filling constraints, it is natural to ask whether the \textit{dipole} filling $\nu_D$,  defined as the dipole moment per unit cell,  constrains the low-energy physics and leads to observable physical consequences.
Note that arguments based on adiabatically threading dipolar flux cannot be directly applied here due to discreteness of the dipole symmetry. 
With PBC, the dipole filling is only defined modulo $1$ due to the exponentiated form of the dipole operator $\hat{D}$ in Eq.~(\ref{eq:dipolegen}),  i.e., its eigenvalues are $e^{i 2 \pi \nu_D}$. 
To clearly delineate the role of dipole filling, we will henceforth work at fixed integer charge filling $\nu \in \mbZ$,  where there are no charge filling constraints,  and show that the answer to the above is affirmative.
%
%%%%%%%%%%%%%%%%%%%%%%%%%%%%%%%%%
%
\paragraph{Background dipole gauge fields:}
To study the constraints imposed by dipole symmetry, we couple the system to background gauge fields on the lattice (see SM~\cite{supmat} for the continuum limit) and consider the effect of a large gauge transformation of the dipole symmetry.
We promote the global symmetry transformation Eq.~\eqref{eq:dipolegen} to a local one: $\hat{b}_j \to  e^{i \alpha_j} \hat{b}_j$.
Under such an arbitrary, site-dependent local phase rotation, the minimal-ranged charge and dipole conserving kinetic term $\hat{b}_{j-1} (\hat{b}_j^\dagger)^2 \hat{b}_{j+1}$ ($\hat{S}_{j-1}^- (\hat{S}_j^+)^2 \hat{S}_{j+1}^-$ for spins with $s \geq 1$) transforms as:
\be
\hat{b}_{j-1} (\hat{b}_j^\dagger)^2 \hat{b}_{j+1} \to e^{-i \left(2 \alpha_j - \alpha_{j+1} - \alpha_{j-1} \right)} \hat{b}_{j-1} (\hat{b}_j^\dagger)^2 \hat{b}_{j+1} \, ,
\label{eq:clustertransform}
\ee
and similarly for its Hermitian conjugate.\footnote{For spinless fermions (or spin-1/2 hard core bosons), the minimal kinetic term has additional sublattice symmetry~\cite{moudgalya2019thermalization}; our results apply once this symmetry is explicitly broken~\cite{supmat}.}

To render the theory gauge invariant, each minimal term is coupled to a single background gauge field $A_j$, leading to the gauge invariant cluster hopping term $e^{-i A_j} \hat{b}_{j-1} (\hat{b}_j^\dagger)^2 \hat{b}_{j+1}$, where $A_j \equiv A_j + 2\pi$ is a compact gauge field that transforms under gauge transformations as $A_j \to A_j  - 2 \alpha_j + \alpha_{j+1} + \alpha_{j-1}$, cancelling the phase shift in Eq.~(\ref{eq:clustertransform}).
For a Hamiltonian that is not simply a sum of minimal terms, the minimal coupling procedure is discussed in the SM~\cite{supmat}.
Since $\alpha_{j+L} - \alpha_j \in 2 \pi \mathbb{Z}$, the Wilson line $\hat{W} = e^{i \phi}$ is gauge-invariant, with $\phi = \sum_j A_j$ the net phase accrued by a dipole around the circle.\footnote{For U$(1)_c$ charge symmetry, $\alpha_j$ is continuous and there is no gauge invariant closed line operator associated with charge hopping. If we Higgs this symmetry from U$(1)$ to $\mathbb{Z}_L$, there is a second gauge invariant line operator $W_D = e^{ i \sum_j j A_j}$~\cite{gorantla2022oned}.}
We will be interested in large gauge transformations which shift $\phi \rightarrow \phi + 2 \pi$.
Phase rotations of the form  $\alpha_j = a_0 + j a_1$ generate the global $\text{U}(1)_c$ (via $a_0$) and $\mathbb{Z}_L^D$ symmetries (via $a_1$) respectively, and do not change the value of the gauge field.
The large gauge transformation thus involves terms quadratic in $j$\footnote{This form ensures that choosing $a_2 \in  (2 \pi /L) \mathbb{Z}$ is sufficient to ensure the periodicity $\alpha_{j+L} = \alpha_j$ mod $2 \pi$.}:
\be
\label{eq:TaylorAlpha}
\alpha_j =  \frac{j(j-L)}{2}a_2 \, , \ \  a_2 \in \frac{2 \pi}{L} \mathbb{Z} \, .
\ee  
Requiring $\alpha_{j+L} = \alpha_j $ (mod $2 \pi$) forces $a_2 \in (2 \pi/L) \mathbb{Z}$.
The gauge transformation Eq.~\eqref{eq:TaylorAlpha} shifts $A_j \rightarrow A_j+ a_2$ (preserving translation invariance of the gauge field), and changes $\delta \phi  =- L a_2 = - 2 \pi$ (for the minimal $a_2 = 2 \pi/L$).  
%
%%%%%%%%%%%%%%%%%%%%%%%%%%%%%%%%%
%
\paragraph{Large gauge transformations and translations:} 
We now study the interplay between large gauge transformations and the generator of translation symmetry. 
Since the operator $\exp(i \sum_j \hat{n}_j \alpha_j)$ generates the phase rotation $\hat{b}_k \to e^{i \alpha_k} \hat{b}_k$, the gauge transformations (\ref{eq:TaylorAlpha}) are generated by the flux insertion operator
\be
\label{eq:lgt}
\hat{O} = \exp \left(i \sum_{j=1}^L \frac{j(j-L)}{2} a_2 \hat{n}_j \right) \, ; \, a_2 = \frac{2\pi m}{L} 
\ee
where for spins, we set $\hat{n}_j \to \hat{S}_j^z$.
Being a large gauge transformation, this operator is unitary, and does not change the many-body spectrum. 
A striking feature of the large gauge transformations Eq.~\eqref{eq:lgt} in dipole conserving models is their non-trivial commutation relation with translation symmetry at general charge and dipole fillings: 
\be
\label{eq:maineq}
\hT \hat{O} = \exp\left(\frac{i \pi m}{L} \left[-2 \hQ_D + \hQ (1 + L) \right] \right) \hat{O} \hat{T} \, ,
\ee
Evaluated on a translation invariant state with fixed integer charge filling~\footnote{Eq.~\eqref{eq:maineq} gives an independent momentum shift at fractional dipole filling even for $\nu \neq \mbZ$ but we restrict to $\nu \in \mbZ$ to highlight the role of dipole symmetry.} $\nu \in \mbZ$, we see that the flux insertion changes the many-body momentum unless
\be \label{eq:maineq2}
- \nu_D + \nu \frac{L+1}{2} \in \mathbb{Z} \ .
\ee
At even integer charge filling $\nu$, this condition requires an integer dipole filling fraction $\nu_D$; at odd integer $\nu$ it requires half-integer (integer) $\nu_D$ when $L$ is even (odd).
Eq. (\ref{eq:maineq2}) constrains the dipole filling fractions at which 1d dipole conserving, translation invariant systems can have gapped, non-degenerate ground states.
To see this for bosonic and integer-spin dipole conserving systems, we fix $\nu$ to be even, and consider a state $\ket{\Psi_P}$ with many-body momentum $P$ i.e., $\hT \ket{\Psi_P} = e^{i P a} \ket{\Psi_P}$ (we restore the lattice constant $a$ here for clarity). Eq.~\eqref{eq:maineq} gives 
\be
\hT \left( \hat{O} \ket{\Psi_P} \right) = e^{-2 \pi i \nu_D} e^{i P a} \left( \hat{O} \ket{\Psi_P} \right) \, .
\ee
The state $\hat{O}\ket{\Psi_P}$ is then an eigenstate of $\hT$ with many-body momentum $Pa - 2 \pi \nu_D$.
Suppose that $\ket{\Psi_0}$ is the ground state of a translation invariant dipole conserving Hamiltonian.
Since $\hat{O}$ is a product of single-site unitary operators (equivalently, a quantum circuit of depth 1), the theorem of Ref.~\cite{gioia2022lre} applies, which states that if there exists a symmetric finite-depth local unitary (FDLU) that boosts a state’s momentum to a different value (mod $2\pi$), then the state is necessarily long-range
entangled.
Hence $\ket{\Psi_0}$ is necessarily symmetry-breaking or gapless when $\nu_D \neq 0$ (mod $1$).
A similar analysis for odd-integer charge fillings shows that the obstruction occurs for any dipole filling $\nu_D \neq 1/2$ ($\nu_D \neq 0$) for $L$ even (odd). 
For instance, in the thin-torus limit of bosonic QH at $\nu = 2$, our constraint rules out Mott insulating states for any $\nu_D \neq 0$ (mod $1$), but allows e.g., charge-density-wave states (consistent with results in Ref.~\cite{nakagawa2017}).

This analysis also applies to fermions~\cite{supmat}, though for systems with one fermionic d.o.f per unit cell, the integer charge filling constraint only permits the empty lattice ($\nu = 0$) or the trivial atomic insulator ($\nu = 1$) as translation invariant gapped ground states, and fractional dipolar fillings cannot be realized.
We extend our analysis to half-integer spin systems in the SM~\cite{supmat}, where we again find that the allowed values of the dipole filling $\nu_D$ at which a symmetric, gapped, and non-degenerate ground state is permitted are specified by Eq.~\eqref{eq:maineq2}. In the SM~\cite{supmat}, we also extend our constraint to systems where the charge is only defined mod $N$. 
Our analysis can also be generalized to systems with higher multipole moment conservation, since the crucial ingredient---a symmetric FDLU that has non-trivial commutation with translations---is furnished by the corresponding uniform large gauge transformation. 
%
%%%%%%%%%%%%%%%%%%%%%%%%%%%%%%%%%
%
\paragraph{Explicit Examples:}
\begin{figure}
    \centering
    \includegraphics[scale=0.9]{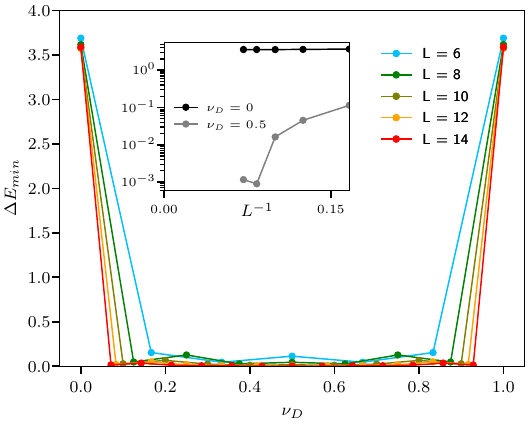}
    \caption{(Color online) Plot of the gap of the spin-1 model Eq.~(\ref{Eq:spin1model}) within the $\nu = 0$ sector as a function of dipole filling $\nu_D$ for various system sizes. 
    Dipole fillings $\nu_D = 0 (\equiv 1)$ show a clear gap, whereas the other fillings appear gapless. 
    The inset shows the gap scaling with inverse system size at dipole fillings $\nu_D = 0$ and $\nu_D = 0.5$, showing evidence for a gap and an absence of a gap respectively.
    Data shown for the Hamiltonian Eq.~(\ref{Eq:spin1model}) with parameters $(J_3, J_4, V_0, V_1, V_2) = (1, 0.3, 1.3, 1.5, -1.2)$.
    }
    \label{fig:gapless}
\end{figure}
We now discuss the implications of the above constraint.
Note first that translation-invariant product states of the form of $\ket{s \cdots s}$, where $s$ denotes the spin, are always gappable and have $\nu = s$ and $\nu_D = s(L + 1)/2\ \text{mod}\ 1$.
For $L$ odd and any integer $s$, or for $L$ even and even integer $s$, $\nu_D = 0$, while for $L$ even and odd integer $s$, we see that $\nu_D = 1/2$, consistent with our results. 
A similar condition can be derived for half-integer spins.
Next, we explore the ground states in the family of interacting spin-1 Hamiltonians  
\begin{align} 
    \hat{H} &= J_3 \sum_j {\left(\hat{S}^-_{j-1} (\hat{S}^+_j)^2 \hat{S}_{j+1}^- + h.c.\right)} \nonumber \\
    &+ J_4 \sum_j{\left(\hat{S}^-_{j-1} \hat{S}^+_j \hat{S}^+_{j+1} \hat{S}^-_{j+2} + h.c.\right)}\nonumber \\
    &+ V_0\sum_j{\hat{S}^z_j} + V_1 \sum_{j}{(\hat{S}^z_j)^2} + V_2\sum_j{\hat{S}^z_j S^z_{j+1}},
\label{Eq:spin1model}
\end{align}
where $\hat{S}^\pm_j$ are spin-1 raising and lowering operators, $\hat{S}^z_j$ measures the spin, and we use PBC.
This model has been extensively studied in the context of Hilbert space fragmentation~\cite{sala2019ergodicity, rakovszky2019statistical, moudgalya2021review, moudgalya2021hilbert}.
Note that since the only states at $\nu = \pm 1$ are the fully polarized states $\ket{\pm\ \cdots\ \pm}$, which are eigenstates of the Hamiltonian (and do satisfy the dipolar constraint), it is not meaningful to discuss a gap in these sectors. 
It remains to show that there are no gappable states at non-zero dipole fillings when $\nu = 0$. 
Numerical explorations of this model for various Hamiltonian parameters suggest that it is gapless within the $\nu = 0$ sector when $\nu_D \neq 0$. Fig.~\ref{fig:gapless} shows the gap as a function of $\nu_D$ for a representative set of parameters, and also illustrates that $\nu_D = 0$ is gappable.
To analytically support these results, we  consider Eq.~\eqref{Eq:spin1model} when $J_4 = 0$, where this Hamiltonian exhibits strong Hilbert space fragmentation and allows explicit solutions within certain Krylov subspaces.
At $\nu = 0$, the Hilbert space sector with $\nu_D = 1/2$ contains a closed Krylov subspace generated by the state $\ket{+\ -\ +\ -\ \cdots\ +\ -}$ for even $L$.
This is the largest Krylov subspace for $\hat{H}$ and contains the ground state for a range of coupling strengths~\cite{rakovszky2019statistical}.  
In the SM~\cite{supmat}, we show that the Hamiltonian within this subspace is a spin-$1/2$ $U(1)$ conserving model (e.g., XX or XXZ) with coupling parameters specified by the parameters of $\hat{H}$.
In this description, the obstruction to constructing a unique, symmetric, and gapped ground state at $(\nu, \nu_D) = (0, 1/2)$ stems from the conventional LSM theorem, which forbids such a state for spin-1/2 models. This restriction applies to \textit{all} models that preserve the above Krylov subspace [when the latter contains the ground state, which we numerically observe to be the case across a large portion of the parameter space of Eq.~(\ref{Eq:spin1model})],  which can be systematically derived~\cite{moudgalya2021hilbert}. 
Hence this solvable limit is also consistent with the filling constraints.   
Similar models and arguments can be written down for higher integer spin systems, as we discuss in \cite{supmat}. 
A second approach to searching for additional gapped states is to use Matrix Product States (MPS). 
In \cite{supmat}, we present constraints that an injective MPS -- i.e., an MPS that can be realized as unique ground states of a local parent Hamiltonian -- with charge and dipole symmetry must satisfy.
We show that these constraints cannot be satisfied by an MPS of bond dimension $2$ at fillings that fail to satisfy Eq.~(\ref{eq:maineq2}), providing further evidence supporting our result.
%
%%%%%%%%%%%%%%%%%%%%%%%%%%%%%%%%%
%
\paragraph{Discussion:} 
We have established dipole-filling enforced constraints for 1d translation invariant dipole conserving systems which prohibit the existence of symmetric, gapped, and non-degenerate ground states except when the dipole filling takes certain special values.
Our analysis highlights the importance of lattice-scale effects for symmetries (such as dipole symmetry) that intertwine non-trivially with translations, leading to differences in the allowed large gauge transformations on the lattice and in the continuum~\cite{supmat}.
Crucially, the constraints obtained by first working on a finite chain and then considering the limit $L \to \infty$ differ from those obtained by analyzing the infinite-volume limit directly: as we have shown here, the former accurately predicts a symmetric, gapped, and unique ground state at $\nu_D = 1/2$ (for certain values of $\nu$ and $L$), whereas the latter ignores the subtlety with boundary conditions and incorrectly prohibits such a state at any fractional dipole filling~\cite{prem2020lsm}.
Similar subtleties with taking the thermodynamic limit in the presence of translation symmetry were recently observed in Ref.~\cite{seiberg2022lsm} and, as our work highlights, become more pronounced in the presence of multipole symmetries.
Our results invite further study of higher-moment conserving systems in higher dimensions and on arbitrary crystalline lattices~\cite{bulmash2023arbitrary}, where their interplay with spatial symmetries will likely lead to additional filling-enforced constraints.
It would also be interesting to study how our filling constraints relate to anomalies for dipole symmetry and to identify the 2d dipole SPTs that host the anomalous 1d theories on their boundaries.
%
%%%%%%%%%%%%%%%%%%%%%%%%%%%%%%%%%
%
\paragraph{Acknowledgements: }
We are particularly grateful to John Sous for discussions at the beginning of this project. 
We also thank Pranay Gorantla, Jonah Herzog-Arbeitman, Michael Knap, Naren Manjunath, Nati Seiberg, and Sahand Seifnashri for stimulating discussions and feedback. This material is based upon work supported by the U.S. Department of Energy, Office of Science, Office of High Energy Physics under Award Number DE-SC0009988 (A.P.), and NSF DMR 1928166 (F. J. B.). A.P. thanks the KITP, which is supported by the National Science Foundation under Grant No. NSF PHY-1748958, for its hospitality during the ``Topology, Symmetry, and Interactions in Crystals" program, when part of this work was completed. 
This work was also supported by the Walter Burke Institute for Theoretical Physics at Caltech and the Institute for Quantum Information and Matter (S.M.).
S.M. also acknowledges the hospitality of the Center for Quantum Phenomena (CQP) at NYU, the Aspen Center for Physics (ACP), and NORDITA Stockholm, where parts of this work were completed. 
ACP is supported by National Science Foundation grants PHY-1607611 and PHY-2210452.
% 
%%%%%%%%%%%%%%%%%%%%%%%%%%%%%%%%%

%---------------------------------------
%BIBLIOGRAPHY
%---------------------------------------

\bibliography{library}

%---------------------------------------
%SUPPLEMENTAL MATERIAL
%---------------------------------------

\onecolumngrid 
\clearpage
\makeatletter 
\begin{center}   
	\textbf{\large Supplementary Material for ``Filling constraints on translation invariant dipole conserving systems"}\\
	[1em]
	Fiona J. Burnell$^1$, Sanjay Moudgalya$^{2,  3,4,5}$, and Abhinav Prem$^{6}$ \\[.1cm]
	{\itshape \small ${}^1$School of Physics and Astronomy, University of Minnesota, Minneapolis, Minnesota 55455, USA \\ 
	${}^2$Department of Physics and Institute for Quantum Information and Matter,\\
California Institute of Technology, Pasadena, California 91125, USA\\
    ${}^3$Walter Burke Institute for Theoretical Physics, California Institute of Technology, Pasadena, California 91125, USA\\
    ${}^4$Department of Physics, Technische Universit\"{a}t M\"{u}nchen (TUM), James-Franck-Str. 1, 85748 Garching, Germany\\
    ${}^5$Munich Center for Quantum Science and Technology (MCQST), Schellingstr. 4, 80799 M\"{u}nchen, Germany\\
    ${}^6$School of Natural Sciences, Institute for Advanced Study, Princeton, New Jersey 08540, USA}\\
	(Dated: \today)\\[1cm]
\thispagestyle{titlepage} 
\end{center} 	
\setcounter{equation}{0} 
\setcounter{figure}{0} 
\setcounter{table}{0} 
\setcounter{page}{1} 
\setcounter{section}{0} 
\renewcommand{\theequation}{S\arabic{equation}} 
\renewcommand{\thesection}{S.\Roman{section}}
\renewcommand{\thetable}{S\arabic{table}} 
\renewcommand\thefigure{S\arabic{figure}} 
\renewcommand{\theHtable}{Supplement.\thetable} 
\renewcommand{\theHfigure}{Supplement.\thefigure}

This supplementary material contains details on the following:
\begin{enumerate}
\item The origin dependence of the dipole filling constraint
\item The minimal coupling procedure for dipole-conserving Hamiltonians
\item Fermionic and half-integer spin systems with dipole conservation
\item Continuum limits of dipole conserving lattice models
\item Mapping of certain spin-1 dipole conserving models onto spin-$\half$ models.
\item Charge and Dipole filling constraints from Matrix Product States
\item Constraints for $\mbZ_N$ Charge and Dipole conserving systems
\end{enumerate}

%%%%%%%%%%%%%%%%%%%%%%%%%%%%%%%%%
%%%%%%%%%%%%%%%%%%%%%%%%%%%%%%%%%

\section{A. Origin dependence of the dipole filling constraint}
\label{sec:origin} 
Here, we detail how the origin dependence of the dipole moment affects the constraints derived in the main text on fillings at which trivial gapped ground states are possible. 
Under a shift of the origin from $ 0$ to $j_o$, we have $\hQ_D = \sum_j{j \hat{n}_j} \to \sum_j{(j-j_o) \hat{n}_j} = \hQ_D - j_o \hQ$, hence the dipole operator transforms as $\hD \to \hD e^{- i \frac{2 \pi j_o}{L} \hQ}$.  
The total dipole filling transforms as $\nu_D \to \nu_D - \nu j_o$; for $\nu \in \mbZ$, $\nu_D$ is unchanged for any integer shift of the origin.  
Shifting the origin also does not affect the commutation relation between $\hT$ and $\hD$, since $\hT$ and $\hQ$ commute.   However, the operator generating a large gauge transformation becomes:
\begin{eqnarray}
\hat{O}' &=& \text{exp} \left[ i \sum_{j=1}^{L}  \frac{ (j-j_0)(j - j_0 - L) }{2} \frac{2 \pi  m}{L}  \hat{n}_{j} \right ]
= 
\text{exp} \left[ i \frac{2 \pi  m}{L}  \left(  \sum_{j=1}^{L}  \frac{ j(j  - L) }{2}  \hat{n}_{j} -  j_0 \sum_{j=1}^{L}  j \hat{n}_{j} +  \frac{j_0 (j_0 + L)}{2} \sum_{j=1}^{L}   \hat{n}_{j} \right ) \right] \nonumber  \\
&=& \hat{O} \hat{D}^{- m j_0} \exp\left({ i \frac{\pi m }{L} j_0(j_0 +L) \hQ}\right)\ .
\end{eqnarray} 
Because $\hD$ does not commute with translations, the commutator between $\hat{O} $ and $\hT$ is modified as follows:
\be
\hT \hat{O} = \exp\left(\frac{i \pi m}{L} \left[-2(\hQ_D- j_0 \hQ)  + \hQ (1 + L) \right] \right) \hat{O} \hat{T} \, ,
\ee
which has an extra factor of $\exp(2 \pi i m j_0 \hQ/L)$ compared to the un-shifted commutator.
Now, on a state with fixed charge filling $\nu$, this results in an excess phase factor of $\exp\left(2 \pi i m j_0 \nu \right)$, which for integer $j_0$ is trivial when $\nu \in \mbZ$ for any value of $m$.
Thus, the constraint on the dipole filling remains invariant for any choice of $j_o \in \mbZ_L$, which is the conventional choice made in the literature.

As discussed in Ref.~\cite{zhang2022pol}, fractional translations correspond to fractional gauge transformations of the translation gauge field, under which we do not expect the dipole filling to remain invariant. 
Nevertheless,  the fact that there is a single value of the dipole filling that allow a symmetric, gapped, and non-degenerate ground state is independent of the choice of origin within the unit cell.
%
%
%%%%%%%%%%%%%%%%%%%%%%%%%%%%%%%%%

\section{B. Minimal Coupling for Dipole-Conserving Hamiltonians}
\label{sec:gauging}
In this section, we show how a generic charge and dipole conserving Hamiltonian $\hat{H}$ is coupled to background gauge fields on the lattice.
We follow the usual minimal coupling procedure used elsewhere in the literature for several types of symmetries (see e.g. Ref.~\cite{shirley2019foliated}.)
In the case of regular symmetries, such as U$(1)_c$ charge conservation, \textit{all} symmetric terms can be expressed as polynomials of the ``minimal" local terms, which generate a bond algebra~\cite{moudgalya2022from}.
Hence a prescription for gauging the minimal local terms uniquely determines the prescription for gauging any symmetric term. 
The situation is subtly different with the dipole symmetry: while not all dipole symmetric terms can be expressed as polynomials of the minimal local terms (this is also related to the presence of Hilbert space fragmentation in such systems~\cite{moudgalya2021hilbert}), they can be generated from the minimal local terms \textit{up to} some extra gauge-invariant operators. In the following, we show that the minimal coupling prescription works well for charge and dipole conserving Hamiltonians in spite of this subtlety. 
We first introduce a gauge field $A_j$ for the charge and dipole symmetries.
Under gauge transformations, this transforms as 
\be
\label{eq:sgt}
A_j \to A_j - 2\alpha_j + \alpha_{j+1} + \alpha_{j-1} \, ,
\ee
with $A_j \equiv A_j + 2\pi$, while the matter degrees of freedom transform as $\hat{b}_j \to e^{i \alpha_j} \hat{b}_j$.
We emphasise that the form of the gauge transformation is imposed by the symmetry alone and not by details of any Hamiltonian.
The ``minimal" symmetric local term that is not simply composed of on-site terms that are invariant under local symmetry transformations is given by $\hh_{m,j} = \hh^+_{m,j} + \hh^-_{m,j}$, where $\hh^+_{m,j} = \hat{b}_{j-1} (\hat{b}_j^\dagger)^2 \hat{b}_{j+1}$ and $\hh^-_{m,j} = (\hh^+_{m,j})^\dagger$.
As stated in the main text, this term can be made gauge invariant by taking $\hh^\pm_{m,j} \to (\hh^\pm_{m,j})^{(g)} = \hat{b}_{j-1} (\hat{b}_j^\dagger)^2 \hat{b}_{j+1} e^{\mp i A_j}$.
Hence the gauge-invariant minimal term is $\hh^{(g)}_{m,j} = (\hh^+_{m,j})^{(g)} + (\hh^-_{m,j})^{(g)}$. 
Now, consider a generic charge and dipole conserving Hermitian term $h_j$ that is not the minimal term $\hh_{m,j}$. 
Splitting as $\hh_j = \hh^+_j + \hh^-_j$, where $\hh^-_j = (h^+_j)^\dagger$, we show that these terms can be expressed as polynomials of the minimal terms as
\be
\hh^+_j G(\{I_k\}) = F\left( \{I_k\}, \{\hh^+_{m,k}\}, \{\hh^-_{m,k}\} \right) \, ,
\label{eq:minimalexp}
\ee
where the terms $I_k$ involve operators such as $\hat{n}_k$ that are invariant under local symmetry transformations, and $F(\cdots)$ and $G(\cdots)$ are  polynomials.
By gauging both sides of the equation, we can then directly deduce the transformation of $\hh^+_j \rightarrow (\hh^+_j)^{(g)}$.
As an example, for a spin-$1$ model term $\hh_j = \hh^+_j + \hh^-_j$, where $\hh^+_j = \hat{S}^-_{j-1}\hat{S}^+_{j}\hat{S}^+_{j+1}\hat{S}^-_{j+2} + h.c.$, we find the following relation 
\be
\left( \hat{S}^-_{j-1} \hat{S}^+_j \hat{S}^+_{j+1} \hat{S}^-_{j+2} \right) [2 + \hat{S}^z_j(1-\hat{S}^z_{j})][2 - \hat{S}^z_j(1+\hat{S}^z_{j})]  = \hh^+_{m,j} \hh^+_{m,j+1}
\ee
showing that a four-site charge and dipole conserving term is proportional to a product of two minimal terms up to gauge invariant operators.
(Here we set $\hat{b}_j \to \hat{S}_j^-$).
From this, we can infer that the gauge-invariant terms can be constructed as 
\be
\hh^+_j \to (\hh^+_j)^{(g)} = \left( \hat{S}^-_{j-1} \hat{S}^+_j \hat{S}^+_{j+1} \hat{S}^-_{j+2} \right) e^{-i \left(A_j + A_{j+1} \right)},\;\;\hh_j \to \hh^{(g)}_j = (\hh^+_j)^{(g)} + (\hh^-_j)^{(g)}
\ee
The above procedure provides a general prescription for coupling any charge and dipole conserving Hamiltonian to a background gauge field $A_j$ that transforms as Eq.~\eqref{eq:sgt} under gauge transformations.
For instance, consider the most general two-body term that conserves these symmetries and has support over some finite range,
\be
\hh_j = \hat{b}_j^\dagger \hat{b}_{j+k} \hat{b}_{j+k+r} \hat{b}_{j+2k+r}^\dagger + \text{h.c.} \, ,
\ee
where $k \geq 1$ and $r \geq 0$. Then, the corresponding gauge-invariant term is given by
\be
\label{eq:generalgt}
\hh_j \to \hat{b}_j^\dagger \hat{b}_{j+k} \hat{b}_{j+k+r} \hat{b}_{j+2k+r}^\dagger \exp\left(i \sum_{p=0}^{k-1} p A_{j+p} + i \sum_{p=0}^r k A_{j+k+p} + i \sum_{p=1}^k (k-p) A_{j+k+r+p} \right) + \text{h.c.} \, .
\ee
One can verify by explicit calculation that such a term is gauge-invariant.
Any higher-order terms can also be made gauge-invariant in a similar way, although we do not attempt to write out their explicit expression.
Note that this discussion applies also to systems with fermionic and spin-$1/2$ degrees of freedom i.e., we can make the term $\hat{c}_j^\dagger \hat{c}_{j+k} \hat{c}_{j+k+r} \hat{c}_{j+2k+r}^\dagger + \text{h.c.}$ gauge-invariant by coupling to background gauge fields in precisely the same manner as specified by Eq.~\eqref{eq:generalgt}.

Note that for the minimal pair-hopping model, the gauge invariant open-line operators
\begin{align}
\hat{d}_{j,j+n} &= \hat{b}_{j+n} \hat{b}^\dag_{j+n+1} e^{i \sum_{k=j}^{j+n} A_k}   \hat{b}^{\dag}_{j-1} b_j \, ,\n
\hat{c}_{j,j+n} &= (\hat{b}^\dag_{j+n+1}\hat{b}_{j+n})^n \hat{b}_{j + n} e^{i \sum_{k=j}^{j+n} (k-j) A_{k}  }  \hat{b}^\dag_{j}
\end{align}
are associated with hopping a dipole from site $j$ to $j+n$, and hopping a charge from site $j$ to $j+n$ while simultaneously creating a compensating dipole, respectively (the latter is only possible when the on-site Hilbert space contains states with charge $\geq n$).

%%%%%%%%%%%%%%%%%%%%%%%%%%%%%%%%%

\section{C. Fermionic and half-integer spin systems with dipole conservation}
Here, we discuss in more detail two classes of dipole-conserving systems that we did not treat in the main text: spin models with half-integer spin, for which we find that the constraints on which fillings can lead to non-degenerate gapped ground states are modified, and charge models where the microscopic degrees of freedom are (spinless) fermions, for which the minimal dipole-conserving term is forbidden by Pauli exclusion.  

\subsection{Constraints on gappability for half-integer spin systems}
\label{sec:half}
As noted in the main text, for a system composed of spin-$s$ degrees of freedom (with a single site per unit cell), we define the dipole operator as
\be
\hQ_D = \sum_{j=1}^L j \hat{S}_j^z  \, .
\ee
Let us now consider the commutator between the translation operator $\hT$ and the dipole operator $\hD$ for half-integer spins:
\be
\hT \hD = \exp\left(2 \pi i \left[ \hat{S}_1^z - \frac{1}{L}\sum_{j=1}^L \hat{S}_j^z\right]\right) \hD \hT \, .
\ee
Note the appearance of the extra term $2 \pi i \hat{S_1^z}$ in the overall phase factor, which has no effect for integer spin systems but modifies the constraint for half-integer spins.
Specifically, for any translation invariant eigenstate $\ket{\phi}$, we find that
\be
\hD \hT \ket{\phi} = \exp \left(2 \pi i \left[S_1^z - \nu \right] \right) \hT \hD \ket{\phi} \, ,
\ee
where $\nu = \frac{1}{L} \sum_{j=1}^L S_j^z$ is the filling.
We hence find a modified constraint in this case, namely that each eigenstate of a dipole conserving translation invariant Hamiltonian with half-integer spin degrees of freedom is degenerate \textit{unless} $\nu \in \mbZ + \half$. For all other fillings, the entire spectrum is necessarily degenerate.
This re-scaling is consistent with the fact that half-integer filling of the spins maps to integer filling of the number of particles, when a Hilbert space of half-integral spin-$s$ is viewed as a Hilbert space of particle configurations with a maximum allowed occupancy of $2s$ per site.

Following the analysis in the main text, we now fix the filling $\nu$ to be half-integer and consider the commutation relation between $\hT$ and the operator that inserts dipole flux $\hat{O}$:
\be
\hat{O} = \exp \left(i \sum_j \frac{j(j-L)}{2} \, a_2 \hat{S}_j^z \right) \, ; \quad \quad a_2 \in \frac{2\pi}{L}
%\begin{cases}
\mbZ,
\ee
We then find that 
\be
\hT \hat{O} = \exp \left(i a_2 \sum_{j=1}^L \left(-j + \frac{L+1}{2} \right) \hat{S}_j^z \right) \hat{O} \hT  \, .
\ee
Let us now consider a state $\ket{\Psi_P}$ with fixed many-body momentum $P$ i.e., $\hT \ket{\Psi_P} = e^{iPa} \ket{\Psi_P}$. Further, let us assume that this state belongs to the sector with half-integer filling $\nu = n + 1/2$ ($n \in \mathbb{Z}_{\geq 0}$). Evaluating the above leads to 
\be
\hT \left(\hat{O} \ket{\Psi} \right) =  \exp\left(i \pi m \left[-2\nu_D + (L+1)\left(n + \frac{1}{2}\right) \right] \right) e^{i P a} \left(\hat{O} \ket{\Psi_P} \right) \, .
\ee
From this relation, we find that the momentum remains unchanged if
\be
\hT \left(\hat{O} \ket{\Psi} \right) = e^{i P a} \left(\hat{O} \ket{\Psi_P} \right) \iff \nu_D = 
\begin{cases}
0 \, , & L = 4k + 3 \text{ and any } n  \\
\frac{1}{4} \, , & L = 4 k \text{ and even } n \text{ or } L = 4k+2 \text{ and odd } n \\
\half \, , & L = 4k+1 \, \text{and any } n \\
\frac{3}{4} \, , & L = 4 k \text{ and odd } n \text{ or } L = 4k+2 \text{ and even } n
\end{cases}
\ee
for $k \in \mbZ$.
Thus, we find that the allowed values of the dipole filling $\nu_D$ at which a symmetric, gapped, and non-degenerate ground state is permitted depend both on the system size as well as the charge filling.
For any other charge and dipole fillings, there exists a finite depth local unitary circuit that changes the many-body momentum of the ground state $\ket{\Psi_0}$, which is hence symmetry-breaking or gapless by the theorem of Ref.~\cite{gioia2022lre}. 
We note that for spin-$\half$ d.o.fs, the charge filling $\nu$ completely constrains the allowed translation invariant gapped ground states and the dipole constraint is trivial in this case.
In other words, to satisfy the  constraint $\nu \in \mbZ + 1/2$, the only translation invariant states that are permitted are $\ket{\uparrow \uparrow \dots \uparrow \uparrow}$ or $\ket{\downarrow \downarrow \dots \downarrow \downarrow}$, with no further freedom left to constrain.
Nevertheless in general, for spin-$s$ d.o.f's, one needs to work in the sector with $\nu \neq \pm s$ in order for the dipole filling to impose a non-trivial constraint, and it is in these sectors that one can obtain distinct weak dipole SPT phases when the system is gapped.

%%%%%%%%%%%%%%%%%%%%%%%%%%%%%%%%%

\subsection{Remarks on fermionic or spin-$1/2$ dipole conserving theories}
For any system with an on-site Hilbert space dimension $\dim(\mathcal{H}_j) \geq 3$, the minimal dipole conserving term that is not simply the product of on-site symmetric terms is the 3-site term considered in the main text. 
However, for spinless fermionic degrees of freedom, this term trivially vanishes and the minimal non-trivial hopping term is the 4-site term
\be
\label{eq:min4site}
\hh_j = \hh^{+}_j + \hh^{-}_j  \ , \ \ \hh^{+}_j  =  \hat{c}_j^\dagger \hat{c}_{j+3}^\dagger \hat{c}_{j+1} \hat{c}_{j+2}  \, ,
\ee
with $\hh^{-}_j = \left( \hh^{+}_j  \right)^\dag$.  The expressions for spin-$1/2$ degrees of freedom are analogous, with $\hat{c}_j \to \hat{S}_j^z$.
However, besides charge-conservation, dipole, and translation symmetries, for even system size $L$ the Hamiltonian $\hat{H} = \sum_j \hh_j$ has additional sub-lattice particle number conservation, since~\cite{moudgalya2019thermalization}
\be \label{Eq:SublatticeSymmetry}
\hat{n}_e = \sum_{j=1}^{L/2}\hat{n}_{2j} \, , \quad \hat{n}_o = \sum_{j=1}^{L/2} \hat{n}_{2j-1} \, ,
\ee
commute with the Hamiltonian $\hat{H} = \sum_j \hh_j$. 
As we will see later, these extra symmetries alter the nature of the underlying gauge theory.

We begin by analyzing generic fermionic systems, where this sublattice symmetry will be broken by longer-ranged dipole-conserving couplings, and showing that the results derived in the main text still hold in this case.
We therefore couple the fermionic theory to the background gauge field $A_j$ using the minimal coupling prescription described above.
As emphasized in that discussion, the gauge transformation should be fixed independently of the details of any Hamiltonian, and since the absence of the three-site term stems from the small Hilbert space dimension and is not fundamentally related to the dipole symmetry, we should still require that $A_j$ transforms as Eq.~\eqref{eq:sgt} under gauge transformations. Then, from Eq.~\eqref{eq:generalgt} we find that the term $\hh_j$ is made gauge-invariant by taking
\be \label{FermionMinimal_1}
\hh_j \to  \hh^{(g)}_j = \left( \hat{c}_j^\dagger \hat{c}_{j+3}^\dagger \hat{c}_{j+1} \hat{c}_{j+2} \right) e^{i\left(A_j + A_{j+1} \right)} + \text{h.c.} \, .
\ee
All other charge and dipole conserving terms can similarly be made gauge-invariant. Crucially, with this coupling, we can include terms in the Hamiltonian which preserve charge and dipole symmetries, but break the sub-lattice symmetry, and couple them consistently to the gauge field $A_j$.

In the generic case, our results from the main text then extend directly to fermionic systems, i.e., we can only obtain symmetric, gapped, and non-degenerate ground states at integer charge filling and dipole filling $0$ or $1/2$ depending on the system size $L$ and charge filling $\nu$.
The only states compatible with integer charge filling in the fermionic case (with one site per unit cell) are a state with no fermions ($\nu = 0$) or a trivial atomic insulator ($\nu = 1$).
The former always has even integer charge filling, and $\nu_D = 0$.
The latter has odd integer charge filling, and $\nu_D = 0$ ($1/2$) when $L$ is odd (even), consistent with our analysis.
\subsubsection{Gauging models with sub-lattice symmetry}
We now turn to the case where we allow {\it only} sublattice-symmetric terms in the Hamiltonian, such that the conservation laws of Eq.~(\ref{Eq:SublatticeSymmetry}) hold. 
While sublattice symmetry is not a property of generic fermionic dipole-conserving systems, it does arise naturally as the lowest order term in the thin-torus limit of the fractional quantum Hall effect~\cite{moudgalya2020quantumhall}, as well as in systems subjected to strong electric fields~\cite{schulz2019stark,taylor2020experimental,  moudgalya2019thermalization}, and thus merits some discussion.
In the presence of sublattice symmetry, $\hh_j$ in Eq. (\ref{eq:min4site}) is the minimal term consistent with all of the symmetries, and is therefore the focus of our minimal coupling prescription.  
As in the main text, we begin by promoting the global symmetry transformation Eq.~(1) to a local transformation $\hat{c}_j \to e^{i \alpha_j} \hat{c}_j$. Under this arbitrary, site-dependent $\text{U}(1)$ transformation, the minimal term in Eq.~\eqref{eq:min4site} transforms as
\be
(\hat{c}^\dag_{j} \hat{c}^\dag_{j+3} \hat{c}_{j+1} \hat{c}_{j+2} + \text{ h.c.}) \to \left( e^{ -i ( \alpha_j + \alpha_{j+3} - \alpha_{j+1} - \alpha_{j+2} )} \hat{c}^\dag_{j} \hat{c}^\dag_{j+3} \hat{c}_{j+1} \hat{c}_{j+2} + \text{ h.c.} \right) \, .
\ee
To render this theory invariant under such a transformation, we define a gauge field $\TA_j$ associated with each cluster, leading to the gauge-invariant cluster hopping terms:
\begin{equation}
    \hat{h}_j \to \hat{h}_j^{(g)} = \left( \hat{c}^\dag_{j} \hat{c}^\dag_{j+3} \hat{c}_{j+1} \hat{c}_{j+2} e^{  i  \TA_j  } + \text{ h.c.} \right) \, , 
\end{equation}
where $\TA_{j} \equiv \TA_{j }+ 2 \pi$ is a compact gauge field that satisfies the gauge transformation rule:
\begin{equation} 
\label{Eq:LatticeGT}
\TA_j \rightarrow \TA_j + \alpha_j + \alpha_{j+3} - \alpha_{j+1} - \alpha_{j+2} \ .
\end{equation}
We emphasize that $\TA_j$ should be viewed as the fundamental gauge field only when charges are individually conserved on the even and odd sublattices (see Eq. (\ref{Eq:SublatticeSymmetry})). Indeed, terms that do not respect the sublattice symmetry cannot be rendered gauge invariant using $\TA_j$. When sublattice symmetry is not present (due, for example, to the presence of longer-ranged pair hopping terms in the Hamiltonian, or because the system size $L$ is odd), we should instead use the gauging procedure described in Eq. (\ref{FermionMinimal_1}), which is achieved by writing 
$\TA_j = A_j + A_{j+1}$.  This is clearly compatible with the gauge transformation (\ref{Eq:LatticeGT}) of $\TA_j$. 
\subsubsection{Gauge invariant line operators in the presence of sublattice symmetry}
To see how the $\TA_j$ gauge theory differs from its $A_j$ counterpart, consider the gauge invariant line operators associated with the gauge field $\TA$.  
As for $A$, the Wilson line $W = e^{ i \phi_{\text{tot}}}$ is gauge invariant, where: 
\begin{equation}
\phi_{\text{tot}}= \sum_{j=1}^L \TA_j
\ .
\end{equation}

The translation invariant large gauge transformations are of the same form as in Eq.~\eqref{eq:TaylorAlpha} in the main text, and imposing PBC $\alpha_{j+L} = \alpha_j \mod 2\pi$ again forces $a_2 \in \frac{2 \pi}{L} \mathbb{Z}$. These shift 
\be \label{Eq:TaGauge}
\TA_j = \alpha_{j} + \alpha_{j+3} - \alpha_{j+2} - \alpha_{j + 1} = 2 a_2 \, .
\ee 
Taking the minimal choice of allowed by PBC, $a_2 = \frac{ 2 \pi }{L}$, thus generates a uniform gauge field $\TA_{j} = \frac{4\pi}{L} \mod 2\pi$ on all bonds, changing 
\begin{equation} \label{Eq:taFluxes}
\delta \phi_{\text{tot}} = \delta \left( \sum_{j=1}^L \TA_j \right )  = 4 \pi \, ,
\end{equation}
or twice the value obtained for the $A$ theory in the main text.  
In particular, the large gauge transformation (\ref{eq:TaylorAlpha}) does not correspond to the minimal flux insertion when $L$ is odd, when there is no sublattice symmetry. 

In the absence of sublattice symmetry, the factors of $2$ appearing in Eqs (\ref{Eq:TaGauge}) and (\ref{Eq:taFluxes}) have an obvious interpretation.  Writing $\TA_j = A_j + A_{j+1}$, we find that for a given $j$, the gauge transformation (\ref{eq:TaylorAlpha}) shifts both $A_j$ and $A_{j+1}$ by $a_2$.
Applying such a gauge transformation to the whole system, we see that  a uniform shift in $A$ shifts each $\TA_j$ by a total of $\frac{ 4 \pi}{L}$, rather than by the minimal amount of $\frac{2 \pi}{L}$ compatible with $\mathbb{Z}_L$ dipole symmetry.
The minimal flux insertion can be obtained by requiring that $\TA$ mod $2 \pi$ be single-valued, while taking the gauge parameter $\alpha$ to obey anti-periodic boundary conditions.
This is always the correct interpretation when $L$ is odd.  
For even system size, when sublattice symmetry is present in the Hamiltonian, $\phi_{\text{tot}}$ can be decomposed into the sum of two distinct line operators $\phi_e$ and $\phi_o$, which can be verified to be gauge-invariant, using the fact that $\alpha_j = \alpha_{j+L} \mod 2\pi$:
\be
\phi_{\text{tot}} = \phi_e + \phi_o \ , \ \ \phi_e = \sum_{j=1}^{L/2} \TA_{2j} \, , \quad \phi_o = \sum_{j=1}^{L/2} \TA_{2j-1} \, .
\ee
(For $L$ odd there is only one closed line operator $\phi_{\text{tot}} \equiv \phi_e + \phi_o$).
The large gauge transformations that shift their values can be obtained by writing:
\be
\alpha_{2j+1} = a_1 j + \frac{a_2}{2}j (j-L) \ , \ \ \   
\alpha_{2j} = b_1 j + \frac{b_2}{2}j (j-L) \ .
\label{eq:alphasublattice}
\ee
Taking $b_1 = a_1, b_2= a_2$ leaves $\phi_e$ unchanged, but shifts $\TA_{2j-1}$ by $a_2$, and thus $\phi_o$ by $\frac{a_2 L}{2}$.
Similarly, taking $b_2 = a_2, b_1 = a_1 - a_2$ leaves $\phi_o$ unchanged, but shifts $\TA_{2j}$ by $a_2$, and thus $\phi_e$ by $\frac{a_2 L}{2}$.   
Thus when $L$ even, using sublattice-dependent gauge transformations we can independently insert $2\pi$ flux on the even and odd sub-lattices.

The existence of two types of Wilson line in this case is a direct consequence of the sublattice symmetry.
First, because the charge is separately conserved on each sublattice, we find two flux insertion operators  $\sum_{j} \alpha_{2j} \hat{n}_{2j}$ and $\sum_j \alpha_{2j+1} \hat{n}_{2j+1}$, with $\alpha_{2j}$ and $\alpha_{2j+1}$ shown in Eq.~(\ref{eq:alphasublattice}) which as noted above we can use to tune the gauge flux separately on the even and odd sublattices.
Moreover with sublattice symmetry, we find two types of dipoles at the lattice scale:
even dipoles $\hat{c}^\dag_{2j -1 } \hat{c}_{2j} $ and odd dipoles $\hat{c}^\dag_{2j } \hat{c}_{2j+1} $. 
Processes hopping a dipole between even and odd sublattices are forbidden by the charge conservation of each sublattice. 
Thus for each dipole type, there is a gauge-invariant open line operator
\be 
\hat{c}^\dag_{2k +1} \hat{c}_{2k} e^{ i  \sum_{j=i}^{k-1} \TA_{2j}} \hat{c}_{2i+1}\hat{c}^\dag_{2i} \ , \ \ 
\hat{c}^\dag_{2k } \hat{c}_{2k-1} e^{ i  \sum_{j=i}^{k-1} \TA_{2j-1}} \hat{c}_{2i}\hat{c}^\dag_{2i-1}  
\ee
associated with the hopping of dipoles on the even and odd sublattices, respectively. 
Thus, models with sublattice symmetry admit a different lattice-scale gauge theory than that described in the main text, with an additional gauge-invariant line operator for even-length systems.  The large gauge transformations shift the flux of each of these operators by $2 \pi$ (and hence the flux of their sum by $4\pi$). However, this difference does not affect the filling constraints: since the operator that carries out a translation-invariant large gauge transformation is unaffected by this distinction, the constraint on the charge and dipole fillings is still given by Eq.~\eqref{eq:maineq} in the main text.  If we do not require  the two sublattices to be related by translation symmetry, additional gapped states are possible since the filling on the two sublattices is then  independent.

%%%%%%%%%%%%%%%%%%%%%%%%%%%%%%%%%

\section{D. Continuum limits of dipole conserving lattice models}
\label{sec:continuum}
Here, we investigate how our lattice models relate to continuum dipolar gauge theories, which have been investigated for example in Ref. \cite{gorantla2022oned}.
For finite-sized dipole conserving systems, taking the continuum limit on the circle is not benign, since it changes the translation symmetry group from $\mathbb{Z}_L$ to U$(1)$ (see Ref.~\cite{seiberg2022lsm}) on the subtleties involved with taking the thermodynamic limit for translation symmetry), and thus the dipolar symmetry from a discrete $\mathbb{Z}_L$ symmetry to a continuous U$(1)$ symmetry.
In the following, we discuss which of our gauge-invariant operators survive this change, and the implications for our classification.
We also describe the open line operators of the lattice theory alluded to in the main text in more detail.

We begin by examining the gauge transformations in the continuum limit.
As the lattice spacing $a \rightarrow 0$, smooth lattice gauge transformations of the form Eq.~\eqref{eq:sgt} are well approximated by:
\be
-2 \alpha_j + \alpha_{j+1} + \alpha_{j-1}\approx  a^2 \partial_x^2 \alpha_j 
\ee
Thus $A_j$ should be viewed as a rank-2 field: defining the  continuum gauge field at position $x_j$ as
\begin{equation}
\label{eq:AxxCt}
A_{xx} (x_j) = \frac{1}{ a^2} A_j, 
\end{equation}
the gauge  transformations Eq.~\eqref{eq:sgt} imply that $A_{xx}$ satisfies the rank-2 gauge transformation rule
\begin{equation}
    A_{xx} \rightarrow A_{xx} +  \partial_x^2 \alpha \ .
\label{eq:Axxtransformation}
\end{equation}
Alternatively, we can start with a continuum rank 2 gauge field transforming according to (\ref{eq:Axxtransformation}), and define the lattice gauge field 
\be \label{eq:ContToLattice}
A_j = \int_{j a - a/2}^{ j a + a/2} d x' \int_{x' - a/2}^{ x' + a/2}d x\  A_{xx}(x) 
\ee
which transforms under the appropriate lattice gauge transformations.

Next, we turn to gauge invariant line operators.
Here, we focus on the spatial line operators, which are related to the flux insertion processes used in the main text to constrain the charge and dipole fillings at which gapped states occur.  
First, we have the closed line operator and its gauge transformation:
\be \label{Eq:WilsonCont1}
\oint A_{xx}\ dx,\;\;\;\delta \left (  \oint A_{xx} dx \right ) = ( \partial_x \alpha (L) - \partial_x \alpha(0) ).
\ee
When $\partial_x \alpha$ is single-valued,  this operator is gauge invariant even without being exponentiated\cite{gorantla2022oned}.
(Exponentiation is typically necessary to ensure that Wilson lines are invariant under large gauge transformations).  
On the lattice, we explicitly considered large gauge transformations for which $\alpha_{j+1} - \alpha_j$ has non-trivial winding, under which these line operators transform non-trivially.  However, 
when $x$ is a continuous variable, such transformations are incompatible with periodic boundary conditions for $\alpha$.  To see this, observe that taking  
\be \label{Eq:ContalphaWind}
\alpha = a_0 + a_1 x + \half a_2 x(x-L)
\ee
yields 
\be
\alpha(x+L) - \alpha(x) = L (a_1 + a_2 x ) \ ,
\ee
and  we can impose $\alpha(x+L) = \alpha(x)$ mod $2 \pi$ at all $x$ only if $a_2 = 0$.  In other words,  requiring invariance under a continuous translation group restricts us to large gauge transformations for which $\partial_x \alpha$ is single-valued.

As pointed out by Ref. \cite{gorantla2022oned}, a physically helpful exponentiated variant of this operator is: 
\be
e^{ i \oint  (\int_{x'- a/2}^{x' + a/2} A_{xx}(x,t)  dx) d x'  } \ .
\ee
This is the continuum analog of the lattice operator $e^{i \sum_j A_j}$ that follows from Eq. (\ref{eq:ContToLattice}).  
There is also an associated gauge invariant open line operator:
\be
b^\dag_{x_2} b_{x_2+a} e^{i  \int_{x_1}^{x_2} d x' \int_{x'}^{x'+a}  d x    A_{xx}(x,t)   } b^\dag_{x_1+a} b_{x_1} 
\ee
associated with pairs of dipoles of length $a$.  This corresponds exactly to the  open line operator associated with dipole hopping on the lattice described in the main text. 
As for the lattice, we also find a second open line operator
\be \label{Eq:ChargeHopAp}
 (b^\dag_{(j+n +1)a} b_{(j+)n a} )^{n} b_{(j + n) a} e^{i \sum_{k=j}^{j+n} (k-j) \int_{j a - a/2}^{ j a + a/2} d x' \int_{x' - a/2}^{ x' + a/2}  A_{xx}(x) dx }  b^\dag_{ja} 
\ee 
associated with moving a charge by a distance $na$, while creating a compensating dipole; this operator is  invariant  only under a discrete subgroup of the continuous translations.  

If the charge symmetry is Higgsed to a discrete, $\mathbb{Z}_{L/a}$ symmetry ($L/a \in \mathbb{Z})$, in addition to the closed line operator $\phi_D = \sum_j j \int_{j a - a/2}^{ j a + a/2} d x' \int_{x' - a/2}^{ x' + a/2}  A_{xx}(x) dx$ suggested by Eq. (\ref{Eq:ChargeHopAp}), we find a second line operator
\be \label{Eq:WilsonCont2}
W_D= e^{ i \phi_D} \ , \ \ \phi_{D} =\frac{1}{a} \oint x \int_{x}^{x+a} d x' A_{xx} (x') \ . 
\ee
which is invariant under both gauge transformations and continuous translations.
Under gauge transformations, it transform as:
\be
\frac{1}{a} \delta \left(  \oint x\int_{x}^{x+a} d x' A_{xx} (x')  \right ) = \frac{L}{a}  \left (\alpha(L+a) - \alpha(L) \right ) \, .
\ee
Thus if the symmetry has been explicitly broken from U$(1)$ to $\mathbb{Z}_{L/a}$, such that  $\alpha(x) = \frac{2 \pi m a}{L}$, $W_D$ is gauge invariant and respects the continuous translation symmetry.  
The lattice version of these closed line operators is:
\be
W_D= e^{ i \phi_D} \ , \ \ \phi_{D} = \sum_j j A_j \ , 
\ee
which is gauge invariant provided we restrict $(\alpha_{j+1} -\alpha_j )  \in \frac{2 \pi }{L} \mathbb{Z}$, and $\alpha_{j+L} = \alpha_j \mod 2\pi$.  

Finally, we briefly comment on how our classification is modified for continuum theories. 
First, the global dipole transformation (which does not affect the winding of any of the line operators discussed above) has the form
\be
\hat{D} = \text{ exp}  \left [ i  \beta  \oint x \rho (x) d x\right] \ .
\ee
 This operator carries out the phase rotation by $\alpha(x) = \beta x$.
We have
\be
\hat{T}_d \hat{D}  =\text{exp}\left [ -i \beta d    \oint  \rho (x) d x \right ]\hat{D} \hat{T}_d
 = \text{exp}\left [ - i \beta d    Q  \right ]\hat{D} \hat{T}_d,
\ee 
where $\hat{T}_d$ is the operator that causes translation by distance $d$.
Requiring $\alpha(x+L) = \alpha(x)$ implies that $\beta = 2 \pi m/L$, $m \in \mathbb{Z}$, so the net phase is $2 \pi d \nu$, where $\nu = Q/L$ is the charge filling.  
Thus if $\hat{T}_d |\psi\rangle = e^{ i p d} |\psi\rangle$, the flux insertion takes 
\be
p \rightarrow p + 2 \pi \nu \ .
\ee
If $d$ can be any real number, then $p$ is not periodic, and we conclude that there are no trivially gapped ground states unless $\nu = 0$.  If we restrict to discrete translations $d = j a$, however, $p$ is periodic modulo $2 \pi /a$, and we recover the expected result that
the system can have a trivial symmetric gapped ground state only if $\nu \in \mathbb{Z}$.  

Our dipole filling constraint arose from considering processes that change the winding of the lattice analogs of operators (\ref{Eq:WilsonCont1}) and (\ref{Eq:WilsonCont2}).  However, as noted above, a uniform gauge field $A_{xx} = \frac{2 \pi}{L a}$ cannot be achieved by any choice of $\alpha$ compatible with periodic boundary conditions.  Instead, we must choose a phase rotation such as
\be
\alpha (x)  = \frac{2 \pi}{L}  \sum_j (x- j a) \Theta(x- j a) \, ,
\ee
where $\Theta(x)$ is the Heaviside function on a circle of circumference $L$, and $L/a \in \mathbb{Z}$.  This gives a gauge field 
\be
A_{xx}(x) =\frac{2 \pi}{L}  \sum_{j} \delta(x - j a) 
\ee
and thus a net winding for $A_{xx}$.  However, this background gauge configuration is invariant only under a discrete subgroup of the original translations, and thus we cannot probe the effect of dipole flux insertions on momentum without implicitly introducing a lattice.  Thus, as for the charged case, our method of finding constraints on dipole fillings compatible with a gapped ground state is valid only for systems with discrete translation symmetry.   

%%%%%%%%%%%%%%%%%%%%%%%%%%%%%%%%%
%%%%%%%%%%%%%%%%%%%%%%%%%%%%%%%%%

%
\section{E. Mapping of certain spin-1 dipole conserving models onto spin-$\half$ models}
In this section, we show that the ground state subspace of certain spin-$1$ dipole conserving models maps onto effective spin-$\half$ models, from which we can directly infer an obstruction to gappability consistent with the results established in the main text.
This mapping is equivalent to one discussed  in~\cite{rakovszky2019statistical} for particular spin-$1$ Hamiltonians, and is also closely related to the mappings in~\cite{moudgalya2019thermalization} for spin-$\half$ charge and dipole conserving models.
We start with the Hamiltonian 
\be
H_3 = J_3 \sum_j {\left(\hat{S}^-_{j-1} (\hat{S}^+_j)^2 \hat{S}_{j+1}^- + h.c.\right)} \,
\ee
with PBC and even system size, and consider the Krylov $\mathcal{K}$ subspace generated by the state $\ket{\psi_0}$, where
\begin{equation}
    \ket{\psi_0} = \ket{+ - + - \cdots + -},\;\;\;\mathcal{K} = \text{span}\{\ket{\psi_0}, H_3 \ket{\psi_0}, (H_3)^2 \ket{\psi_0}, \cdots\}. 
\label{eq:initstate}
\end{equation}
Since $H_3$ conserves charge and dipole moment, this entire subspace is composed of states with charge and dipole filling factors $(\nu, \nu_D) = (0, \half)$. 
However, as a consequence of the Hilbert space fragmentation of $H_3$~\cite{sala2019ergodicity, rakovszky2019statistical, moudgalya2021review, moudgalya2021hilbert}, $\mathcal{K}$ is not the full subspace of states with $(\nu, \nu_D) = (0, \half)$. 
Multiple actions of $H_3$ connect $\ket{\psi_0}$ to a subset of states with the same charge and dipole moment, in particular those of the form 
\begin{equation}
    \ket{0 \cdots 0 + 0 \cdots 0 - 0 \cdots 0 \cdots + 0 \cdots 0 - 0 \cdots 0},
\label{eq:stringorder}
\end{equation}
which form a basis for the Krylov subspace $\mathcal{K}$.
States of this form are said to possess ``string order"~\cite{affleck1988valence, rakovszky2019statistical, moudgalya2021hilbert}, which means that they have an alternating pattern of $+$'s and $-$'s once the $0$'s are ignored. 
This Krylov subspace by definition is closed under the action of $H_3$, and is numerically observed to be the largest Krylov subspace~\cite{sala2019ergodicity, rakovszky2019statistical}.  It also contains the full ground state of $H_3$.

We now show that the action of $H_3$ has a nice form under a transformation from the original lattice, where spins live on sites labelled by $\{j\}$, to the dual lattice, where the spins live on the links labelled by $\{j + \half\}$.
We perform the following mapping:
\begin{equation}
    \ket{+}_{j} \leftrightarrow \ket{\downarrow\uparrow}_{j - \half, j+\half},\;\;\;
    \ket{-}_{j,j+1} \leftrightarrow \ket{\uparrow\downarrow}_{j + \half},\;\;\;\ket{0}_j \leftrightarrow \ket{\sigma\sigma}_{j - \half, j+\half},\;\;\sigma \in \{\uparrow, \downarrow\},
\end{equation}
where the spin $\sigma$ in the mapping of $0$'s is chosen such that it is consistent with the mappings assigned to $+$'s and the $-$'s.  
This mapping is an isomorphism between the Hilbert space spanned by even $L$ spin-1's with string order, denoted by $\mH_{s}$ and the Hilbert space of $L$ spin-$\half$'s denoted by $\mH_{\half}$.
The Krylov subspace $\mathcal{K}$ is a subspace of $\mH_s$.
Given the isomorphism of these Hilbert spaces, we can also write down a dictionary between the operators in these two spaces:
\begin{align}
\hat{S}^z_{j} & = \hat{\sigma}^z_{j+\half} - \hat{\sigma}^z_{j-\half}, \nonumber \\
\hat{S}^-_{j-1}(\hat{S}^+_{j})^2 \hat{S}^-_{j+1} + \hat{S}^+_{j-1}(\hat{S}^-_{j})^2 \hat{S}^+_{j+1} & =  \half\left(\hat{\sigma}^x_{j-\half}\hat{\sigma}^x_{j+\half} + \hat{\sigma}^y_{j-\half}\hat{\sigma}^y_{j+\half}\right) 
\label{eq:opdict}
\end{align}
where $S^{x,y,z}$ and $\sigma^{x,y,z}$ respectively denote spin-1 and spin-$\half$ operators, and by equality we mean that the action of the L.H.S. within $\mH_{s}$ maps onto the action of the R.H.S. within $\mH_{\half}$.
With this, we see that the spin-1 charge and dipole operators map as
\begin{equation}
    \sum_j{\hat{S}^z_j} = 0,\;\;\;\sum_j{j \hat{S}^z_{j}} = -\sum_j{\hat{\sigma}^z_{j + \half}} + L \hat{\sigma}^z_{\half}.
\label{eq:chargedipoleopmap}
\end{equation}
Any spin-1 state with charge and dipole filling $(\nu, \nu_D)$ hence maps onto spin-$\half$ state with spin filling $\half - \nu_D$.
Since the Krylov subspace $\mathcal{K}$ has charge and dipole filling factors $(\nu, \nu_D) = (0, \half)$, in the spin-$\half$ language we are restricted to the sector with spin filling $0$. 
Using the operator dictionary of Eq.~(\ref{eq:opdict}), it is possible to build a variety of spin-1 dipole conserving models that map onto spin-$\half$ spin conserving models when restricted to the Krylov subspace generated by $\ket{\psi_0}$. 
For example, the pure dipole conserving Hamiltonian $H_3$ studied in \cite{rakovszky2019statistical} maps onto the spin-$\half$ XX model.
In addition, electrostatic interactions that are diagonal in the computational basis [such as those in Eq.~(\ref{Eq:spin1model}) in the main text] can be added to $H_3$ without  altering the fact that the Krylov subspace $\mathcal{K}$ is closed.   These make the effective Hamiltonian within the subspace a spin-$\half$ XXZ model. 
We can then ask if appropriate gapped spin-$\half$ models can be engineered so that the corresponding spin-1 model has a gap, at least within the subspace $\mathcal{H}_s$. 
However, there is a clear obstruction from the usual Lieb-Schultz-Mattis (LSM) theorem~\cite{LSM1961, oshikawa2000commensurability} to constructing gapped spin-$\half$ models at spin filling factors $\nu \notin \mbZ + \half$.
Hence, within the realm of models that possess the closed subspace $\mH_s$ of string ordered states, the obstruction to constructing charge and dipole conserving spin-1 models gapped within the $(\nu, \nu_D) = (0, \half)$ sector maps onto the usual LSM obstruction to constructing unique gapped ground states of spin-$\half$ models. 
One might then wonder if this mapping generalizes to higher spin dipole conserving models, which, within certain Krylov subspaces, might then map onto integer spin models that are gappable. 
In order for this to happen for arbitrary spin-$S$, it can be explicitly checked that the types of dipole conserving terms need to be carefully chosen so that similar closed Krylov subspaces composed of states with string order exist. 
With such appropriately chosen terms, the operator dictionary of Eq.~(\ref{eq:chargedipoleopmap}) remains valid, where the L.H.S. are spin-$S$ operators and the R.H.S. are spin-$\frac{S}{2}$ operators. 
Spin-$S$ states with charge and dipole filling factors $(\nu, \nu_D) = (0, \half)$ then always map onto spin-$\frac{S}{2}$ states with spin filling $(\frac{S}{2} - \nu_D)\ \text{mod}\ 1$. 
If $S$ is odd, the effective model is a half-integer model with integer spin filling, and the usual LSM obstruction still holds. 
If $S$ is even, the effective model is an integer spin model with half-integer charge filling, where again a LSM-type filling constraint holds~\cite{oshikawa2,hastings2004}.
Hence, in these cases with analytical tractability, there are obstructions to gappability, which provide additional evidence for the validity of our claims in the main text. 
\section{F. Charge and dipole filling constraints from Matrix Product States}
In this section, we directly analyze the filling constraints on charge and dipole fillings by studying the action of these symmetries on Matrix Product States (MPS). 
We write down the necessary conditions to  construct a finite-bond dimension injective MPS that is an eigenstate of both charge and dipole symmetries, with any possible eigenvalues. 
Such an injective MPS admits a gapped parent Hamiltonian with the MPS as its unique ground state with periodic boundary conditions~\cite{perezgarcia2007matrix, cirac2021matrix}.
For concreteness, we consider a spin-$S$ system with $d = 2S + 1$ degrees of freedom.
We restrict to integer $S$ for simplicity; the case of half-integer $S$ can be worked out in close analogy.
The unitary operators $U_c$ and $U_d$ that represent U$(1)_c$ charge symmetry and the $\mbZ_L$ dipole symmetry are given by
\begin{equation}
    U_c(\theta) = \bigotimes_{j = 1}^L{u(\theta)},\;\;\;\;\;\;U_{d}(n) = \bigotimes_{j = 1}^L{u\left(\theta = \frac{2\pi n}{L}\right)^j},\;\;\;u(\theta) = e^{i \theta S^z}, 
\label{eq:symops}
\end{equation}
where $S^z$ is the on-site spin-$S$ operator, $\theta \in [0, 2\pi)$, and $0 \leq n < L$.
Any state $\ket{\psi}$ with charge and dipole filling $\nu$ and  $\nu_D$ respectively should satisfy
\begin{equation}
    U_c(\theta) \ket{\psi} = e^{i L \theta \nu}\ket{\psi},\;\;\;U_d(n)\ket{\psi} = e^{i 2\pi n \nu_D}\ket{\psi} \, .
\label{eq:cdeigenvalues}
\end{equation}
For a one dimensional system, a gapped, non-degenerate ground state can be represented to arbitrary precision with an injective MPS.
A translation-invariant MPS with PBC is a state defined as~\cite{cirac2021matrix}
\begin{equation}
    \ket{\psi} = \sum_{\{m_j\}}{}{\ \textrm{Tr}[A^{[m_1]} A^{[m_2]} \cdots A^{[m_L]}]\ket{m_1, m_2, \cdots m_L}},
\label{eq:MPSdefn}
\end{equation}
where $m_j$ labels the state on site $j$, $A^{[m_j]}$ is a $\chi \times \chi$ matrix that encodes the state, where $\chi$ is referred to as the bond dimension of the MPS.
$A$ is viewed as $d \times \chi \times \chi$ tensors, where the $d$-dimensional index is referred to as the physical index, and the $\chi$-dimensional indices are referred to as auxiliary indices or ancilla.
For any MPS symmetric under a global on-site symmetry such as $U_c(\theta)$, the local tensors satsify the condition~\cite{perezgarcia2008string, cirac2021matrix}
\begin{equation}
    \sum_{m'}{u(\theta)_{m, m'} A^{[m']}} = e^{i\phi(\theta)} V(\theta) A^{[m]} V(\theta)^\dagger,
\label{eq:MPSsymmetry}
\end{equation}
where $m, m'$ run over the $d$-dimensional physical index, $V$ is a $\chi$-dimensional matrix and the multiplication on the R.H.S. is over the auxiliary index.
It is easy to see that any MPS with the tensors satisfying Eq.~(\ref{eq:MPSsymmetry}) is an eigenstate of the global symmetry operators $U_c(\theta)$ with eigenvalue $e^{i \phi(\theta)L}$.
With the conditions that $\phi(\theta = 0) = 0$ and those of Eq.~(\ref{eq:cdeigenvalues}) we obtain
\begin{equation}
    \phi(\theta) = \theta \nu.
\label{eq:phiform}
\end{equation}
Setting $\theta = 2\pi$ in Eq.~(\ref{eq:MPSsymmetry}), we get
\begin{equation}
    A^{[m]} = e^{i 2\pi \nu} V(\theta = 2\pi) A^{[m]} V(\theta = 2\pi)^\dagger,
\label{eq:singlematrix}
\end{equation}
where we have used that $u(\theta = 2\pi) = 1$.

Suppose the $\chi$-dimensional MPS $\{A^{[m]}\}$ is injective under blocking $\ell \geq \ell_0$ sites, i.e., we have
\begin{equation}
    \text{span}_{\{m_j\}}\{A^{[m_1]} A^{[m_2]} \cdots A^{[m_\ell]}\} = \mathcal{M}_{\chi \times \chi},
\label{eq:injectivity}
\end{equation}
where $\mathcal{M}_{\chi\times\chi}$ is the full matrix algebra of $\chi \times \chi$ matrices, and $\ell_0$ is the minimum number of sites needed to be blocked to see the injectivity. 
Any tensor $A$ satisfying Eq.~(\ref{eq:injectivity}) for some finite $\ell$ is referred to as a \textit{normal tensor}~\cite{cirac2021matrix}. 
Eq.~(\ref{eq:injectivity}) means that some linear combination of the matrices $\{A^{[m_1]} A^{[m_2]} \cdots A^{[m_\ell]}\}$ is the identity matrix.
Since $V(\theta = 2\pi)$ is unitary, we can also apply Eq.~(\ref{eq:singlematrix}) to this linear combination to show that 
\begin{equation}
    e^{i 2\pi \nu \ell} = 1\;\;\forall\;\;\ell \geq \ell_0.\;\;\implies \nu \in \mathbb{Z}.
\label{eq:chargecondition}
\end{equation}
This recovers the standard LSM-type filling constraint that integer charge filling is a necessary condition to construct a gappable MPS ground state~\cite{oshikawa2000commensurability, hastings2004lieb, chen2011classification, schuch2011classifying, prakash2020elementary}.
We now impose the condition that the MPS should be symmetric under $U_d(n)$, as shown in Eq.~(\ref{eq:cdeigenvalues}).
Acting $u(\theta = \frac{2\pi n}{L})^j$ on the physical index of the $j$'th MPS tensor, using Eqs.~(\ref{eq:MPSsymmetry}) and (\ref{eq:phiform}) we obtain  
\begin{equation}
    \sum_{m'}{\left[u\left(\theta = \frac{2\pi n}{L}\right)\right]^j_{m, m'} A^{[m']}} = e^{i\frac{2\pi n \nu j}{L}} \left[V\left(\theta = \frac{2\pi n}{L}\right)\right]^j A^{[m]} \left[V\left(\theta = \frac{2\pi n}{L}\right)^\dagger\right]^j \, .
\label{eq:MPSsymmetrydipole}
\end{equation}
Hence the action of $U_d(n)$ on the full MPS reads
\begin{equation}
        U_d(n)\ket{\psi} = e^{i\phi_n \frac{L(L+1)}{2}}\sum_{\{m_j\}}{}{\ \textrm{Tr}[V_n A^{[m_1]} V_n A^{[m_2]} V_n \cdots V_n A^{[m_L]}]\ket{m_1, m_2, \cdots m_L}} 
\end{equation}
where we have defined
\begin{equation}
    \phi_n = \phi\left(\theta = \frac{2\pi n}{L}\right) = \frac{2\pi n \nu}{L},\;\;V_n = V\left(\theta = \frac{2\pi n}{L}\right).
\end{equation}
Using Eq.~(\ref{eq:cdeigenvalues}), we obtain
\begin{equation}
        e^{i\phi_n \frac{L(L+1)}{2}}\sum_{\{m_j\}}{}{\ \textrm{Tr}[V_n A^{[m_1]} V_n  \cdots V_n A^{[m_L]}]\ket{m_1 \cdots m_L}} = e^{i 2\pi n \nu_D}\sum_{\{m_j\}}{}{\ \textrm{Tr}[A^{[m_1]} \cdots A^{[m_L]}]\ket{m_1, \cdots m_L}}.
\label{eq:MPSdipoleconstraint}
\end{equation}
Since we want both sides of this equation to be same injective MPS, at the level of the local tensors we should have~\cite{cirac2021matrix}
\begin{equation}
    \sqrt{V_n} A^{[m]} \sqrt{V_n} = e^{i \Upsilon} G A^{[m]} G^{-1}\;\;\forall\;m,\;\;\text{where}\;\;e^{i \Upsilon L} = e^{i 2\pi n (\nu_D - \nu \frac{L+1}{2})}, 
\label{eq:dipolelocaltensor}
\end{equation}
where $G$ is an invertible $\chi \times \chi$ matrix. 
It remains to determine if there are any constraints on $\Upsilon$ so that Eq.~(\ref{eq:dipolelocaltensor}) can be satisfied for some choice of $G$. 
In order to determine this, we analyze the structure of $V_n$ and $A^{[m]}$ coming from the charge symmetry constraint of Eq.~(\ref{eq:MPSsymmetry}). 
It is known that $V(\theta)$ should form a projective representation of the symmetry group of $u(\theta)$~\cite{cirac2021matrix}, which in this case is U$(1)$. 
Using the fact $\nu \in \mbZ$ and Eq.~(\ref{eq:singlematrix}), we can use the injectivity of $A^{[m]}$ to conclude that that $V(\theta = 2\pi) = 1$.
Its general form can be written as
\begin{equation}
    V(\theta) = \text{diag}(e^{i v_1 \theta}, e^{i v_2 \theta}, \cdots, e^{i v_\chi \theta}),\;\;\;v_\mu \in \mbZ.    
\end{equation}
Expanding Eq.~(\ref{eq:MPSsymmetry}) to first order in $\theta$, we then obtain
\begin{gather}
    A^{[m]} + i\theta \sum_{m'}{S^z_{m,m'}A^{[m']}} + \mathcal{O}(\theta^2) =(1 + i \theta \nu + \mathcal{O}(\theta^2))(A^{[m]} + i \theta [M, A^{[m]}] + \mathcal{O}(\theta^2))\nonumber \\
    \implies m A^{[m]} = [M, A^{[m]}] + \nu A^{[m]},
\label{eq:tensorconstraint}
\end{gather}
where $M = \text{diag}(v_1, v_2, \cdots, v_\chi)$. 
Eq.~(\ref{eq:tensorconstraint}) implies some constraints on the structure of the tensors $\{A^{[m]}\}$ for any charge symmetric MPS, which has been studied earlier in great detail~\cite{singh2011tensor}. 
This can in principle be used to constrain the solutions in Eq.~(\ref{eq:dipolelocaltensor}).

While we are not able to obtain a useful constraint in the general case, we focus on $\nu = 0$ in spin-1 systems, and restrict to bond dimension $\chi = 2$, where we can perform a brute-force search for solutions to Eq.~(\ref{eq:dipolelocaltensor}).
Using Eq.~(\ref{eq:tensorconstraint}), we can show that any such MPS of bond dimension $\chi = 2$ can be brought to the form
\begin{equation}
    A^{[+]} = 
    \begin{pmatrix}
    0 & c_+ \\
    0 & 0
    \end{pmatrix},\;\;\;
    A^{[0]} = 
    \begin{pmatrix}
    c_0 & 0 \\
    0 & c'_0
    \end{pmatrix},\;\;\;
    A^{[-]} = 
    \begin{pmatrix}
    0 & 0 \\
    c_- & 0
    \end{pmatrix},
    V(\theta) = 
    \begin{pmatrix}
    e^{i v_1 \theta} & 0 \\
    0 & e^{i v_2 \theta}
    \end{pmatrix}
\label{eq:chi2MPS}
\end{equation}
where $c_+$, $c_-$, $c_0$, and $c'_0$ are some non-zero numbers and $v_1, v_2 \in \mbZ$. 
Performing this search for solutions of Eq.~(\ref{eq:dipolelocaltensor}), we obtain that there are no non-trivial solutions for $G$ and $V_n$ that preserve the injectivity of $A^{[m]}$ unless $\Upsilon = 0$.
This implies that the fillings satisfy
\begin{equation}
    \nu \frac{L+1}{2} = \nu_D\;\;\text{mod}\;\;1. 
\end{equation}
Hence we recover the filling constraint in the main text from this MPS point of view. 
It would be interesting to generalize this argument to MPS of arbitrary finite bond dimensions, where it is likely that results on the block structures of U$(1)$ charge symmetric MPS, derived in \cite{singh2011tensor} would be useful.

\section{G. Constraints for $\mbZ_N$ charge and dipole conserving systems}
In this Section, we consider the case where the charge is only conserved modulo $N$, in which case we obtain distinct results from those considered in the main text. Note that such systems have recently been investigated in the context of dipole-symmetry protected topological phases in one dimension~\cite{han2024,lam2024}.

First, note that if charge is conserved only mod $N$, then the dipole moment can also be conserved \textit{at most } mod $N$.
In particular, this allows single charges to hop by $N$ sites.
If the system size is $L = m N + r$, then with periodic boundary conditions, it follows that by a series of $m$ consecutive hops by $N$ sites, the particle can move a distance $r$ around the circle.
Thus, the dipole moment is conserved at most mod $r$, which reduces the $\mbZ_N$ dipole symmetry to a $\mbZ_{\text{gcd}(L, N)}$ symmetry.
To see this explicitly, let $L = q n$ and $ N = p n$, where $n = \text{gcd}(L, N)$.
Then $q n = m p n + r$,  hence $r$ is also divisible by $n$ and the dipole symmetry is at least $\mbZ_n$.
On the other hand, the charge can be moved by any amount $k N + t r$ mod $L$ for any $k$ and $t$, which is equivalent to $(k - m t) N$ mod $L$.
Since $\text{gcd}(L, N) = n$, there always exist integers $\ell, \ell'$ such that $\ell N = \ell' L + n$; hence choosing $t = 1$, there always exists $k$ such that the expression $(k - m) N$ mod $L = n$.
%
%\sout{By assumption, $q$ and $p$ have no common factors and so there also exists a $t$ for which $r + t N = (q-pm+ t p )n = n$ mod $L$}. {\fb Let's double check, I now don't remember where I got this fact from}.\sanjay{Tried to rewrite}  
%
Hence $k n + t r = n$ mod $L$ for an appropriate choice of $k$ and $t$, and the dipolar symmetry is also at most $\mbZ_n$.
Here, we will restrict ourselves to the case where the periodicity of the dipole symmetry and the charge symmetry are the same, in which case it suffices to consider values of $N$ for which $\text{gcd}(L, N) = N$, i.e., $L/N  \in \mathbb{Z}$.  

Under this reduced charge symmetry, the phases $\alpha_j$ associated with global U(1) phase rotations now take on discrete values,  $\alpha_j \in \frac{2 \pi}{N} \mathbb{Z}$; it is thus natural to impose this restriction when gauging the dipole symmetry as well. When the gauge transformations are discrete, we identify a second gauge-invariant line operator:
\be
W_D = e^{ i \phi_D} \, , \quad \phi_D =  \sum_{j=1}^L j A_j \, .
\ee
%{\fb Do we really need the factor of 2?  Or is it gauge invariant without?}
Under gauge transformations,
\begin{eqnarray}
\label{Eq:ClosedGT2}
\delta(\phi_D) &=   L ( \alpha_{L+1}-\alpha_L ) - (  \alpha_L - \alpha_0 ), 
\end{eqnarray}
and $W_D$ is gauge-invariant under discrete gauge transformations, for which $   L ( \alpha_{L+1}-\alpha_L )  = p N \frac{ 2 \pi m}{N}$ is an integer multiple of $2 \pi$.

As discussed in Sec. D, the corresponding open string operators are associated with moving single charges.
Note that, on the lattice, with $U(1)$ charge conservation such motion is prohibited and neither $W_D$ nor the corresponding open strings are gauge invariant operators. However, when the $U(1)$ charge symmetry is Higgsed to $\mathbb{Z}_N$, we see that $W_D$ is gauge invariant, and $\phi_D$ describes the phase acquired by a charge that has travelled around the circle.  In this case gauge-invariant open string operators also exist, associated with transporting a charge by $N$ sites.  
We emphasize however that the Wilson line operator discussed in the main text is always gauge invariant, irrespective of the status of $W_D$.

The general form of an operator carrying out a large gauge transformation is 
\be
\hat{O} = e^{ i \sum_j \alpha_j n_j} \ .
\ee
Let us consider $\alpha_j$ of the general form
\be
\alpha_j = \frac{a_2}{2} j^2 + a_1 j \ ,
\ee
which is the minimal form required for a large gauge transformation, since $A_j$ transforms with the second lattice derivative of $\alpha$.  
To find the coefficients $a_1$ and $a_2$, we first note that for all $j$ we must have (both for the $\mbZ_N$ case discussed here and for the U(1) case discussed in the main text)
\be
\alpha_{j+L} - \alpha_j \in 2 \pi \mathbb{Z} \implies j L a_2 + L\left(\frac{L}{2} a_2 + a_1 \right) \in 2 \pi \mathbb{Z} \ ,\;\;\implies\;\;a_2 =  \frac{2 \pi n}{L}\;\;\implies   L(\pi n + a_1 ) \in 2 \pi \mathbb{Z}
\ee
where $n$ is some integer.
Thus we see that if $L$ is even, we can choose $a_1$ to be any integer  multiple of $2 \pi /L$ (in the main text, we have specifically taken $a_1 = - n \frac{L}{2} \times \frac{2 \pi}{L}$).
On the other hand, if $L$ is odd, we need $a_1$  to be an odd-integer multiple of $n \frac{\pi}{L}$  (in the main text, we have specifically taken $a_1 =  - L \times \frac{n \pi}{L}$).

When the charge symmetry is $U(1)$, this is the only condition, and we are free to choose $n$ to be any integer, as we did in the main text.
However, when the charge symmetry is a discrete $\mbZ_N$ symmetry we additionally require that 
\be
\alpha_{j+1} -  \alpha_j \in \frac{2 \pi}{N} \mathbb{Z} \implies a_2 j + \left(\frac{a_2}{2}  + a_1\right)  \in \frac{2 \pi}{N} \mathbb{Z} \, .
\ee
This can hold for all $j$ only if 
\be
a_2 =  \frac{2 \pi k}{N}  \ , \ \   (\pi k + N a_1) \in 2 \pi \mathbb{Z} \ 
\ee 
where $k$ is an integer.
In other words, $a_1 = \frac{\pi}{N}(2m + k)$, where m is any integer.  
Thus the gauge parameter has the form:
\be \label{Eq:alphajN}
\alpha_j = a_2 \frac{j (j +1)}{2} + \tilde{a}_1 j \ , \ \ a_2 = \frac{2 \pi k}{N} \ , \ \  \tilde{a}_1 =  \frac{2 \pi m}{N}.
\ee
Now, we can ask whether it’s always possible to choose $k=1$ (i.e.  $a_2 =  \frac{2 \pi}{N}$), without violating periodicity of $\alpha$.  From the discussion above, we see that our choice of $a_1 = \frac{\pi}{N} (2 m +k)$ is compatible with periodicity when $L$ is odd, but not necessarily when $L$ is even.
Specifically, we have
\be \label{Eq:alphaPerC}
\alpha_{j+L} - \alpha_j = 2 \pi  ( k j+ m) \frac{L}{N}  + 2 \pi k \frac{ L}{N}\frac{(L+1)}{2}  
\ee
The first term is always an integer multiple of $2 \pi$, because we have assumed $L$ is divisible by $N$.  
If $L$ is odd, the second term is also an integer multiple of $2 \pi$, irrespective of $k$.
If $L$ is even, and $L/N$ is even, the second term is again an integer multiple of $2\pi$. 
However, if $L$ is even and $L/N$ is odd, the second term is an integer multiple of $2 \pi$ only when $k$ is even.
Thus in this case, the naive minimal choice of $k=1$ is incompatible with both periodic boundary conditions and $\mathbb{Z}_N$ valued gauge transformations.

%{\fb End alternative discussion}.

%Under large gauge transformations {\fb of the first (second) type}, we find  $2 \delta \phi_D =-L(L+1) a_2$ {\fb ($2 \delta \phi_D =-L(2L- (N-1)) a_2$)}.
%
%Further, we find:
%\be
%\label{eq:maineq}
%\hT \hat{O} = \begin{cases}  \exp\left(i \frac{a_2}{2} \left[-2 \hQ_D + \hQ (1 + L) \right] \right) \hat{O} \hat{T} \\ 
%  \exp\left(i \frac{a_2}{2} \left[-2 \hQ_D + \hQ (1 + N) \right] \right) \hat{O} \hat{T} 
 % \end{cases}
%\ee
%Under large gauge transformations, we find {\fb  $2 \delta \phi_D =-L(L+1) a_2$.}
%
We now ask whether these large gauge transformations change the many-body momentum.  We have:
\be
\label{eq:maineq}
\hT \hat{O} =  \exp\left(-i\left [ a_2 \hQ_D + \tilde{a}_1\hQ  \right] \right) \hat{O} \hat{T} 
=    \exp\left(-2 \pi i\left [  k \frac{ \hQ_D}{N} +  m \frac{\hQ}{N}   \right] \right) \hat{O} \hat{T} \, .
\ee
We see that when $Q/L$ is an integer (so that $Q/N$ is also an integer), any choice of $m$ gives an equivalent trivial phase factor.  If $L$ is odd, then $k$ can be any integer.   This gives the constraint that $Q_D= 0 $ modulo $N$, and there is only one value of the dipole moment for which the system can be gapped.

%If $L$ is even, then in the first line $a_2 = 4 \pi m /N$; this allows for $Q_D/N = 1/2$ to yield a gapped ground state. If $N$ is odd, however, the second line gives the same criterion as above, namely that only the state with $Q_D/N \in \mathbb{Z}$ is gapped. In contrast, when $N$ is even both criteria allow for gapped ground states at both $Q_D/N = 0 $ and $1/2$ modulo $1$.  In particular, this is possible when $N = L$. In fact, this agrees with the analysis of the $U(1)$ case considered in the main-text, which found that the dipolar filling for even $L$ was integral (half-integral) when the charge filling was an even (odd) integer. With charge defined only modulo $L$, the distinction between even and odd integer charge filling is no longer possible, and we have two gapped ground states at charge filling $0$ (modulo $1$), distinguished by their dipole filling fraction.

If $L$ is even, and $L/N$ is odd, however, then $k$ must be even; this means that there are two distinct gapped ground states, associated with dipole filling fractions $Q_D/N = 0, 1/2$. In particular, this is possible when $N = L$. In fact, this agrees with the analysis of the $U(1)$ case considered in the main-text, which found that the dipolar filling for even $L$ was integral (half-integral) when the charge filling was an even (odd) integer.  With charge defined only modulo $L$, the distinction between even and odd integer charge filling is no longer possible, and we have two gapped ground states at charge filling $0$ (modulo $1$), distinguished by their dipole filling fraction.

Indeed, if $L/N = p$ is odd (such that $N$ is even, since $L$ is  even by assumption), these gapped ground states correspond to uniform filling with $0$ and $N/2$ particles per site, respectively. A uniformly filled state with charge $m$ on each site has $Q_D =  m L (L+1)/2 =m L  (N-1)/2 $ mod $L$, respectively (this follows from the fact that $p N (p N+1)/2- p N  (N-1)/2 = N(p  + N p (p-1)/2 )$, which is divisible by $N$). If $N$ is odd, then $(N-1)/2$ is an integer, so this is always equivalent to $0$ filling, consistent with the fact that the unique gapped ground state in this case has $Q_D / N \in \mbZ$. If $N$ is even, however, $(N-1)/2$ is a half-integer, and thus when $p$ is odd, $p N (N-1)/2$ is a half-integer multiple of $N$; thus taking $m=1$ we obtain a  gapped ground state with $\nu_D = (N-1)/2$, which is half-integral. Evidently, a second gapped ground state with $Q_D= 0 $ mod $N$ is obtained by taking $m=0$.  

One might ask what happens in the case that $N$ is not a divisor of $L$. As described above, the dipole symmetry becomes reduced to $\mbZ_{n}$ ($n=\text{gcd}(L, N))$, while the charge symmetry remains $\mathbb{Z}_N$, so that $\alpha_j \in \frac{ 2 \pi}{N} \mbZ$. We thus have 
\be
\delta \phi_D \in 2 \pi \frac{L}{N} \mbZ =  2 \pi \frac{q}{p} \mbZ \, ,
\ee
where we have taken $L = q n, N = p n$.  Thus $W_D = \text{exp}[ ip \phi_D]$ is gauge invariant.  If $n=1$ then $p= N$ and $W_D$ is a trivial operator, but if $n>1$, $p < N$ and it represents a physically measurable phase obtained by hopping a single charge $p < N$ times around the system. 

As for the classification, the  large gauge transformations are as described in Eq. (\ref{Eq:alphajN}).  However, the single-valuedness constraint (\ref{Eq:alphaPerC}) requires that
\be \label{}
\alpha_{j+L} - \alpha_j  
= 2 \pi  ( k j+ m) \frac{q}{p}  + 2 \pi k \frac{ q}{p}\frac{(L+1)}{2} \ .
\ee
Since this must hold for all $j$, 
$k$ and $m$ be multiples of $p$ in general to ensure that the first term is an integer multiple of $2 \pi$.  When $L$ is even and $q$ is odd, $k$ must be a multiple of $2 p$.  
The commutator between translation and dipole flux insertion then gives:
\be
\hT \hat{O} =  \exp\left(- 2 \pi i  \frac{p}{N} \left[ l  \hQ_D + s \hQ  \right] \right) \hat{O} \hat{T} 
= \exp\left(- 2 \pi  q \left[l \nu_D + s \nu  \right] \right) \hat{O} \hat{T} 
\ee
where $s = m/p$ is any integer, and $l = k/p $ is an even integer for $L$ even and $q$ odd, and any integer otherwise.  

It is also worth revisiting the possible dipole symmetry transformations in this case. The dipole symmetry operator has the general form
$
D = e^{ i \sum_j a_1 j n_j} 
$.  A $\mathbb{Z}_N$ dipole symmetry would suggest that $a_1 = \frac{ 2 \pi m}{N}$; however, single-valuedness of the symmetry transformation in periodic boundary conditions also requires that $L a_1 \in 2 \pi \mathbb{Z}$.  Thus if $L/N = q/p$, then $a_1 = s p \frac{ 2 \pi}{N}$ and we find that neither dipole symmetry nor dipole flux insertion yield an obstruction to having a gapped ground state with $\nu =Q/L =  1/p$. However, by assumption $L$ is not divisible by $p$; hence this is not possible since $Q$ must be an integer. Indeed, the smallest integer $m$ for which $Q = m L/p \in \mathbb{Z}$ is $m=p$, since (again, by assumption), $L$ and $p$ share no common factors.  In other words, a trivial gapped ground state is in fact possible only at integer charge fillings, consistent with the well-known conclusion in the $U(1)$ case.  
At integer charge filling, the remaining criterion that a gapped ground state must satisfy is that $q l \nu_D \in \mathbb{Z}$. If either $L$ is odd or $q$ is even, $l$ can be any integer and we obtain a single gapped ground state with a trivial dipole moment. If $L$ is even and $q$ is odd, however, then $l$ is even and we expect two distinct gapped ground states.  

Indeed, as above, these distinct gapped ground states can be realized at uniform filling $m$.  In this case, $Q_D = m L (L+1)/2 = m \frac{qn}{2} ( qn + 1)$.  Let us first consider the case $m=1$.  When $n$ is even, this cannot be $0$ modulo $N = n p$; if it were, it would follow that $\frac{q}{2} ( qn + 1) = r p$ for some integer $r$, which is impossible since $\frac{q}{2} ( qn + 1)$ is a half-integer.  It follows that taking $m=1$, we get $\nu_D = (L+1)/2$, which is a gapped ground state with charge filling $\nu = 1$  and half-integer dipole filling (since $L$ is even by assumption).  A distinct gapped ground state can be obtained by taking $m=0$.

%---------------------------------------
%---------------------------------------

\end{document}